\documentclass[10pt,journal,compsoc]{IEEEtran}
\ifCLASSOPTIONcompsoc
  \usepackage[nocompress]{cite}
\else
  \usepackage{cite}
\fi
\ifCLASSINFOpdf
\else
\fi
\usepackage{amsmath}

\usepackage{algorithmic}
\ifCLASSOPTIONcompsoc
 \usepackage[caption=false,font=footnotesize,labelfont=sf,textfont=sf]{subfig}
\else
 \usepackage[caption=false,font=footnotesize]{subfig}
\fi
\usepackage{url}

\usepackage[bookmarks=false]{hyperref}
\usepackage[table]{xcolor}
\usepackage[ruled]{algorithm2e}
\usepackage{amssymb,amsfonts}
\usepackage{balance}
\usepackage{booktabs,threeparttable}
\usepackage{makecell, multirow}
\usepackage{color}
\usepackage{enumerate}
\usepackage{float}
\usepackage{graphicx}
\usepackage{listings}
\usepackage{subfig}
\usepackage{textcomp}
\usepackage{xcolor}
\usepackage{enumitem}
\usepackage{circledsteps}
\usepackage{tikz}
\usepackage{multirow}
\usepackage{multicol}
\usepackage[notext]{stix2}

\floatstyle{plain}
\newfloat{lstfloat}{H}{myc}

\makeatletter
\def\bstctlcite{\@ifnextchar[{\@bstctlcite}{\@bstctlcite[@auxout]}}
\def\@bstctlcite[#1]#2{%
 \@bsphack
 \@for\@citeb:=#2\do{%
 \edef\@citeb{\expandafter\@firstofone\@citeb}%
 \if@filesw\immediate\write\csname #1\endcsname{\string\citation{\@citeb}}\fi}%
 \@esphack}
\makeatother

\makeatletter
\def\lst@makecaption{%
  \def\@captype{table}%
  \@makecaption
}
\makeatother

\makeatletter
\newcommand{\removelatexerror}{\let\@latex@error\@gobble}
\makeatother

\newcommand*\circleddark[2]{\tikz[baseline=(char.base)]{
            \node[shape=circle,scale=0.8, fill=#1,inner sep=1pt, text=white, font=\bfseries] (char) {#2};}}

\def\code#1{\texttt{#1}}
\def\BibTeX{{\rm B\kern-.05em{\sc i\kern-.025em b}\kern-.08em
    T\kern-.1667em\lower.7ex\hbox{E}\kern-.125emX}}

\definecolor{grey}{rgb}{0.3,0.3,0.3}
\definecolor{neonred}{rgb}{1.0,0.09,0.36}
\definecolor{neonblue}{rgb}{0.3,0.3,1.0}
\definecolor{neongreen}{rgb}{0.0,0.67,0.498}
\definecolor{lightneonblue}{rgb}{0.86,0.86,0.98}

\DeclareMathOperator*{\minimize}{minimize}

\DeclareMathOperator*{\argmax}{arg\,max}

\lstdefinestyle{CABI} {
    language=c,
    frame=single,
    mathescape=false,
    basicstyle=\footnotesize\ttfamily,
    keywordstyle=\bfseries\ttfamily,
    morekeywords={size_t},
    breaklines=true,
    columns=fullflexible
}

\begin{document}
\bstctlcite{IEEEexample:BSTcontrol}

%
\title{A Probabilistic Machine Learning Approach to \\ Scheduling Parallel Loops with \\ Bayesian Optimization}
%
%
%
%
\author{Kyurae~Kim,~\IEEEmembership{Student~Member,~IEEE,}
        Youngjae~Kim,~\IEEEmembership{Member,~IEEE,}
        and~Sungyong~Park,~\IEEEmembership{Member,~IEEE}
\IEEEcompsocitemizethanks{\IEEEcompsocthanksitem K. Kim is with the Department of Electronics Engineering, Sogang University, Seoul, Republic of Korea.\protect\\
E-mail: msca8h@sogang.ac.kr
\IEEEcompsocthanksitem Y. Kim and S. Park are with the Department of Computer Science and Engineering, Sogang University, Seoul, Republic of Korea.\protect\\
E-mail: \{youkim, parksy\}@sogang.ac.kr
}
\thanks{Manuscript received June 26, 2020; revised August 21, 2020; accepted September 7 2020.
  This work was supported by the Next-Generation Information Computing Development Program through the National Research Foundation of Korea (NRF) funded by the Ministry of Science, ICT (2017M3C4A7080245).
  This paper was presented in part of the 27th IEEE International Symposium on Modeling, Analysis, and Simulation of Computer and Telecommunication Systems (MASCOTS'19) held in Rennes, France, 2019. \textit{(Corresponding author: Sungyong~Park.)}}
}

\IEEEtitleabstractindextext{%
\begin{abstract}
  This paper proposes \textit{Bayesian optimization augmented factoring self-scheduling} (BO FSS), a new parallel loop scheduling strategy.
  BO FSS is an automatic tuning variant of the factoring self-scheduling (FSS) algorithm and is based on Bayesian optimization (BO), a black-box optimization algorithm.
  Its core idea is to automatically tune the internal parameter of FSS by solving an optimization problem using BO.
  The tuning procedure only requires online execution time measurement of the target loop.
  In order to apply BO, we model the execution time using two Gaussian process (GP) probabilistic machine learning models.
  Notably, we propose a \textit{locality-aware GP} model, which assumes that the temporal locality effect resembles an exponentially decreasing function.
  By accurately modeling the temporal locality effect, our locality-aware GP model accelerates the convergence of BO.
  We implemented BO FSS on the GCC implementation of the \code{OpenMP} standard and evaluated its performance against other scheduling algorithms. 
  Also, to quantify our method's performance variation on different workloads, or \textit{workload-robustness} in our terms, we measure the \textit{minimax regret}. 
  According to the minimax regret, BO FSS shows more consistent performance than other algorithms.
  Within the considered workloads, BO FSS improves the execution time of FSS by as much as 22\% and 5\% on average. 
\end{abstract}

}

\maketitle


\IEEEdisplaynontitleabstractindextext

%
\IEEEpeerreviewmaketitle

\IEEEraisesectionheading{\section{Introduction}\label{sec:introduction}}

\IEEEPARstart{L}{oop} parallelization is the de-facto standard method for performing shared-memory data-parallel computation.
Parallel computing frameworks such as \code{OpenMP}~\cite{dagum_openmp_1998} have enabled the acceleration of advances in many scientific and engineering fields such as astronomical physics~\cite{regier_cataloging_2018}, climate analytics~\cite{kurth_exascale_2018}, and machine learning~\cite{baydin_etalumis_2019}.
A major challenge in enabling efficient loop parallelization is to deal with the inherent imbalance in workloads~\cite{durand_impact_1996}.
Under the presence of load imbalance, some computing units (CU) might end up remaining idle for a long time, wasting computational resources.
It is thus critical to schedule the tasks to CUs efficiently.

Early on, dynamic loop scheduling algorithms~\cite{kruskal_allocating_1985, hummel_factoring_1992, tzen_trapezoid_1993, hagerup_allocating_1997, lucco_dynamic_1992, bast_scheduling_2000, banicescu_load_2002} have emerged to attack the parallel loop scheduling problem.
However, these algorithms exploit a limited amount of information about the workloads, resulting in inconsistent performance~\cite{ciorba_openmp_2018}.
In our terms, they lack \textit{workload-robust} as their performance varies across workloads.

Meanwhile, \textit{workload-aware} scheduling methods have recently emerged.
These methods, including the history-aware self-scheduling (HSS, ~\cite{kejariwal_historyaware_2006}) and bin-packing longest processing time (BinLPT,~\cite{penna_binlpt_2017, penna_comprehensive_2019}) algorithms, utilize the static imbalance information of workloads.
Static imbalance is an imbalance inherent to the workload that is usually caused by algorithmic variations.
Unlike dynamic imbalance, which is caused by the environment of execution, static imbalance can sometimes be accurately estimated before execution.
In these cases, workload-aware methods aim to exploit the static imbalance information for scheduling.
On the other side of the coin, these methods are inapplicable when the static imbalance information, or a workload-profile, is not provided.
In many high-performance computing (HPC) applications, static imbalance is often avoided by design.
Even if such imbalance is present, it is usually unknown unless extensive profiling is performed.
As a result, workload-aware methods can only be applied to a limited range of workloads or challenging to use at best.

As discussed, both dynamic and workload-aware methods have limitations.
Thus, additional efforts must be made to find the algorithm best-suited for a particular workload.
Practitioners often need to try out different scheduling algorithms and manually tune them for the best performance, which is tedious and time-consuming.
To resolve the issues of dynamic and workload-aware scheduling methods, we propose \textit{Bayesian optimization augmented factoring self-scheduling} (BO FSS), a workload-robust parallel loop scheduling algorithm.
BO FSS automatically infers properties of the target loop only using its execution time measurements.
Since BO FSS doesn't rely on a workload-profile, it applies to a wide range of workloads. 

In this paper, we first show that it is possible to achieve robust performance if we are able to appropriately \textit{tune} the internal parameters of a classic scheduling algorithm to each workload individually.
Based on this observation, BO FSS tunes the parameter of \textit{factoring self-scheduling} (FSS,~\cite{hummel_factoring_1992}), a classic dynamic scheduling algorithm, only using execution time measurements of the target loop.
This is achieved by solving an optimization problem using a black-box global optimization algorithm called Bayesian optimization (BO,~\cite{shahriari_taking_2016}).
BO is notable for being data efficient; it requires a minimal number of measurements until convergence~\cite{alipourfard_cherrypick_2017}.
It is also able to efficiently handle the presence of noise in the measurements.
These properties previously led to successful applications such as compiler optimization flag selection~\cite{letham_constrained_2018}, garbage collector tuning~\cite{dalibard_boat_2017}, and cloud configuration selection~\cite{alipourfard_cherrypick_2017}.
Based on these properties of BO, our system is able to improve scheduling efficiency with a minimal number of repeated executions of the target workload.

To apply BO, we need to provide a \textit{surrogate model} that accurately describes the relationship between the scheduling algorithm's parameter and the resulting execution time.
By extending our previous work in~\cite{kim_robust_2019a}, we propose two types of probabilistic machine learning models as surrogates.
First, we model the total execution time contribution of a loop as~\textit{Gaussian processes} (GP).
Second, for workloads where the loops are executed multiple times in a single run, we propose a \textit{locality-aware GP} model.
Based on the assumption that the temporal locality effect resembles exponentially decreasing functions, our locality-aware GP can accurately model the execution time using exponentially decreasing function kernels from~\cite{swersky_freezethaw_2014}.
As a result, it is able to achieves faster convergence of BO when applicable.

We implemented BO FSS as well as other classic scheduling algorithms such as chunk self-scheduling (CSS,~\cite{kruskal_allocating_1985}), FSS~\cite{hummel_factoring_1992}, trapezoid self-scheduling (TSS,~\cite{tzen_trapezoid_1993}), tapering self-scheduling (TAPER,~\cite{lucco_dynamic_1992}) on the GCC implementation~\cite{gcc_gcc_2018} of the \code{OpenMP} parallelism framework.
Then, we evaluate BO FSS against these classical algorithms and workload-aware methods including HSS and BinLPT.
To quantify and compare the robustness of BO FSS, we adopt the \textit{minimax regret} metric~\cite{mcphail_robustness_2018, doi:10.1080/01621459.1951.10500768}.
We selected workloads from the Rodinia 3.1~\cite{che_characterization_2010} and GAP~\cite{beamer_gap_2017} benchmark suites for evaluation.
Results show that our method outperforms other scheduling algorithms by improving the execution time of FSS as much as 22\% and 5\% on average.
In terms of workload-robustness, BO FSS achieves a regret of 22.34, which is the lowest among the considered methods.

\vspace{0.05in}
The key contributions of this paper are as follows:
\begin{itemize}
\item \textbf{We show that, \textit{when appropriately tuned}, FSS can achieve workload-robust performance} (Section~\ref{backgroundrelated}). In contrast, the performance of dynamic scheduling and workload-aware methods varies across workloads.
\vspace{0.05in}
\item \textbf{We apply BO to tune the internal parameter of FSS} (Section~\ref{method}).
Results show that BO FSS achieves consistently good performance across workloads (Section~\ref{eval}).
\vspace{0.05in}
\item \textbf{We propose to model the temporal locality effect of workload using locality-aware GPs} (Section~\ref{labo}).
Our locality-aware GP incorporates the effect of temporal locality using exponentially decreasing function kernels.
\vspace{0.05in}
\item \textbf{We implement BO FSS over the \code{OpenMP} parallel computing framework} (Section~\ref{impl}).
Our implementation includes other classic scheduling algorithms used for the evaluation and is publicly available online\footnote{Source code available in \url{https://github.com/Red-Portal/bosched}}.
\vspace{0.05in}
\item \textbf{We propose to use \textit{minimax regret} for quantifying workload-robustness of scheduling algorithms} (Section~\ref{eval}).
According to the minimax regret criterion, BO FSS shows the most robust performance among considered algorithms.
\end{itemize}


\section{Background and Motivation}\label{backgroundrelated}
  In this section, we start by describing the loop scheduling problem.
  Then, we show that dynamic scheduling and workload-aware methods lack what we call workload-robustness.
  Our analysis is followed by proposing a strategy to solve this problem.

\begin{figure*}
  \centering
  \subfloat[Accuracy of load estimation]{
    \includegraphics[scale=0.4]{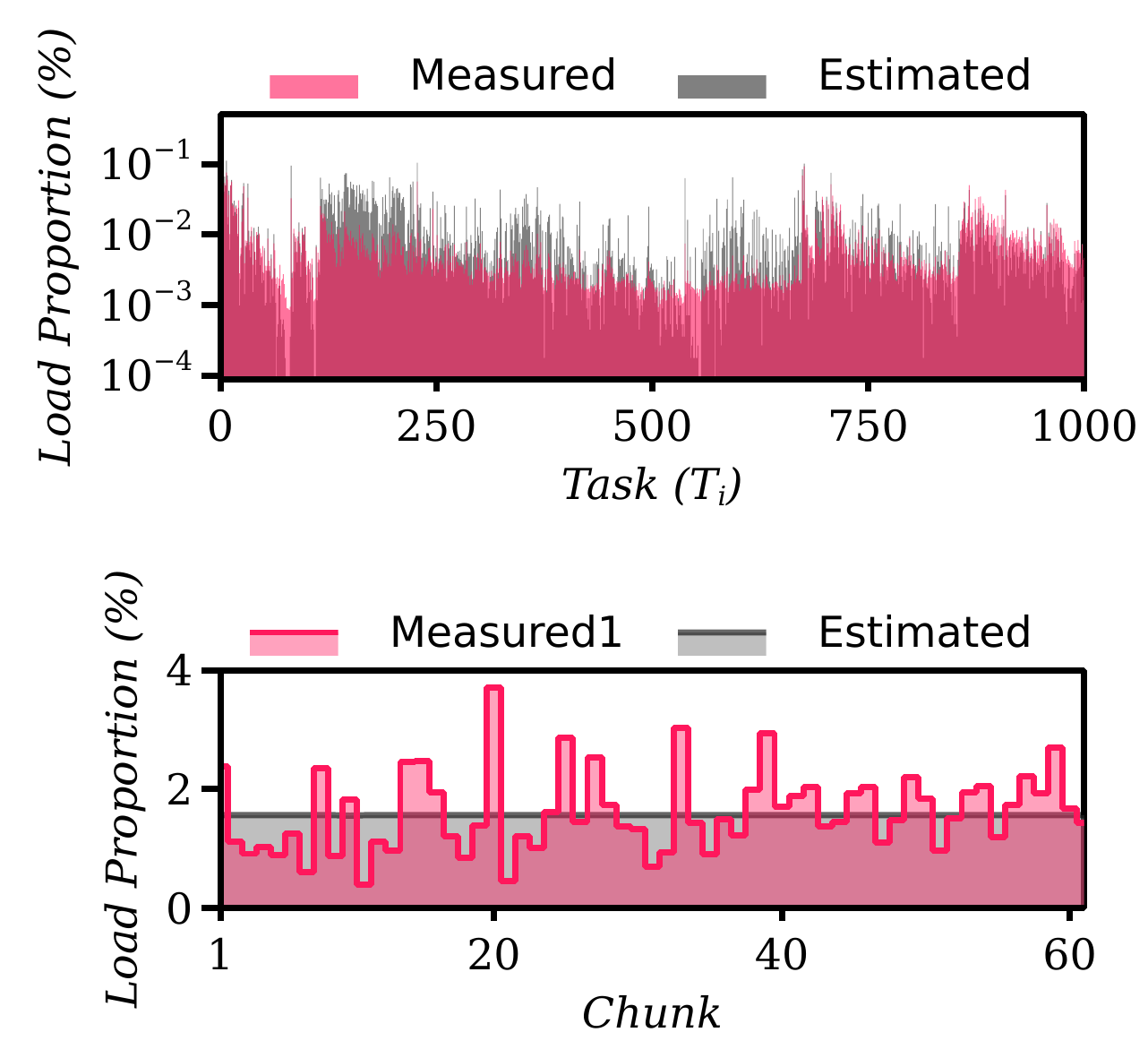}\label{fig:loadprof}
  }
  \subfloat[Low static imbalance]{
    \includegraphics[scale=0.4]{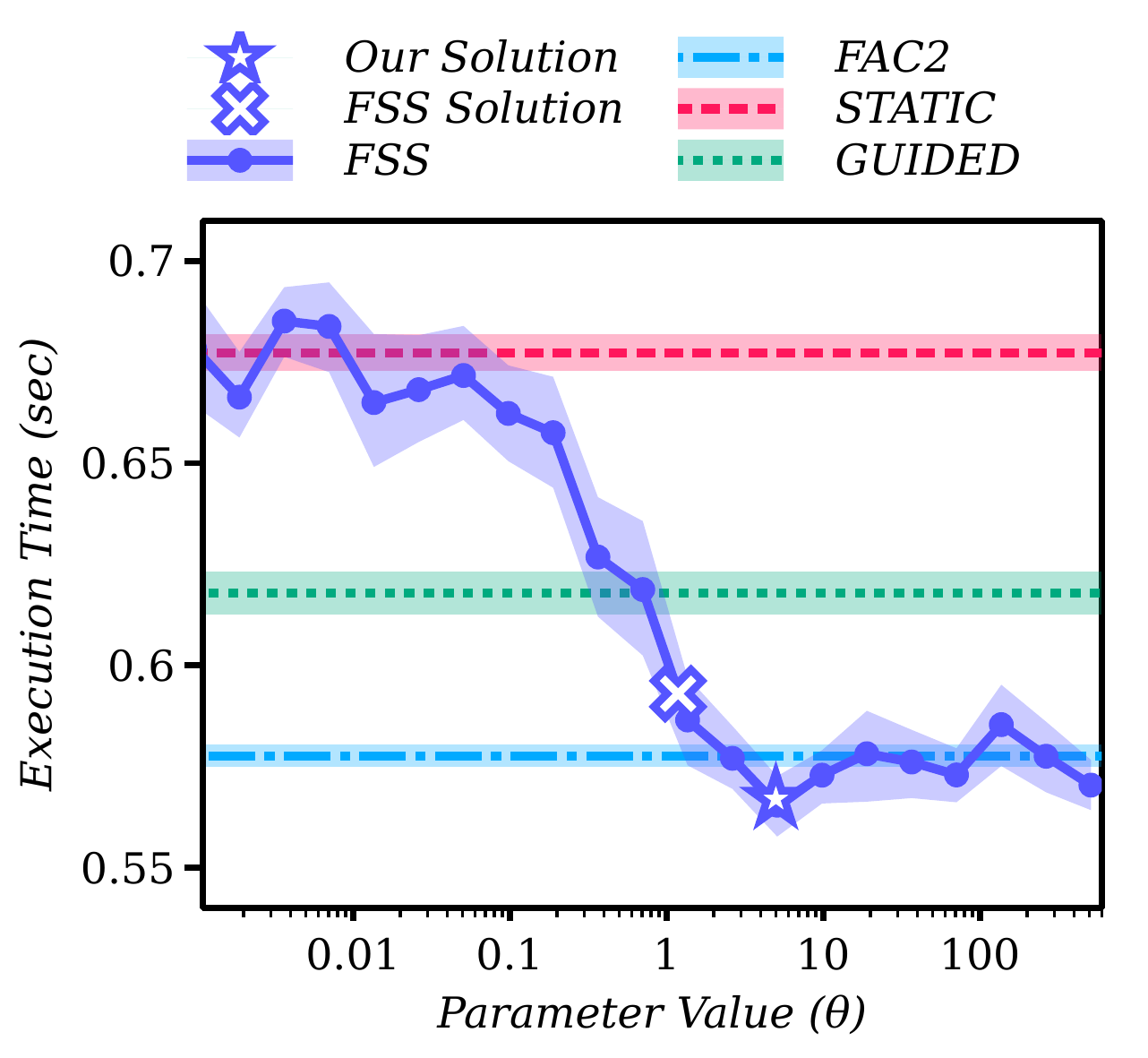}\label{fig:motiv1}
  }
  \subfloat[High static imbalance]{
    \includegraphics[scale=0.4]{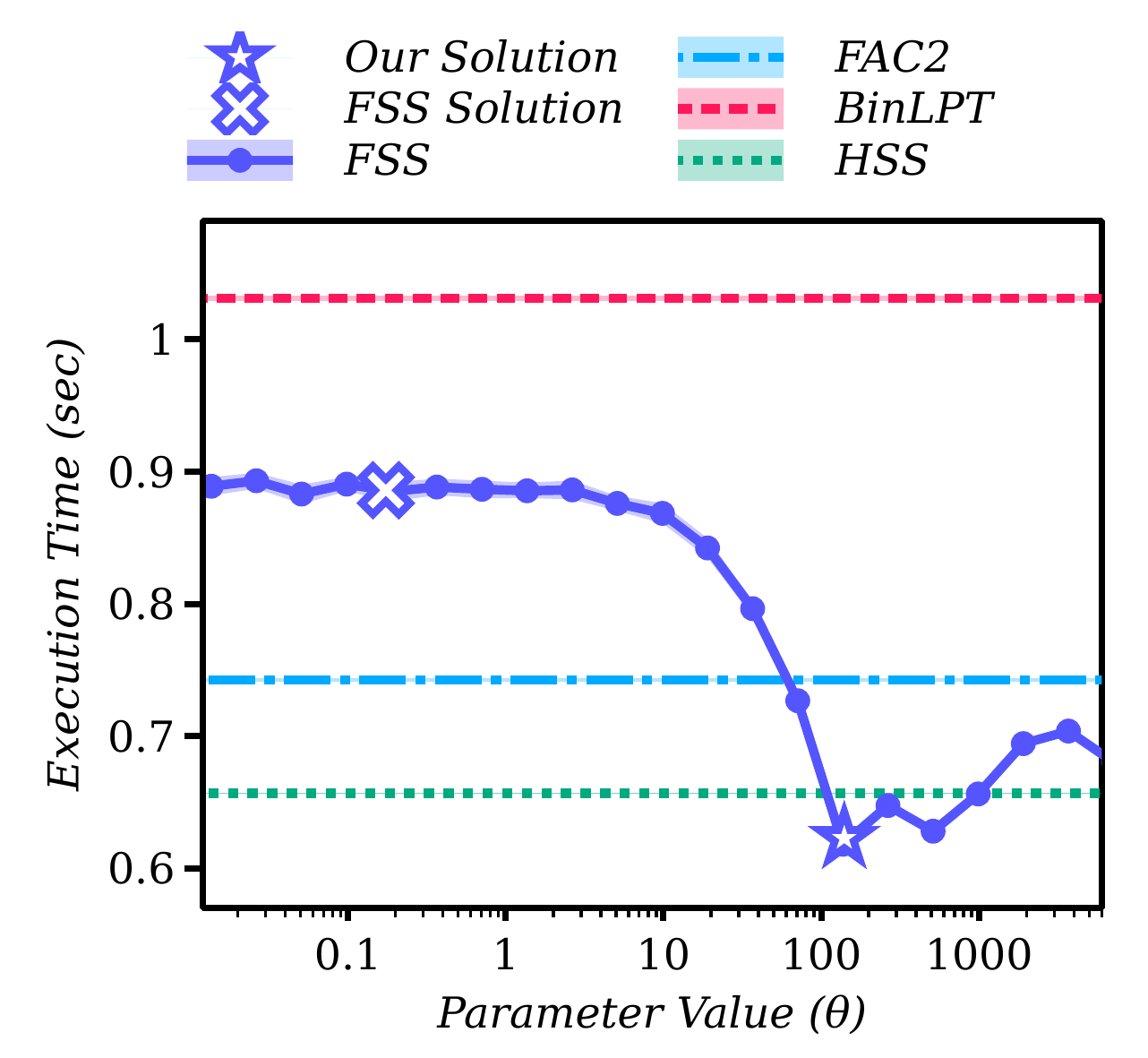}\label{fig:motiv2}
  }
  \caption{((a), top) Discrepancy between the workload-profile and actual execution time of the tasks. ((a), bottom) Discrepancy between the load of the chunks created by BinLPT, and their actual execution time.
    (b-c) Effect of the internal parameter ($\theta$) of FSS on a workload with homogeneous tasks ((b), low static imbalance, \code{lavaMD} workload) and a workload with non-homogeneous tasks ((c), high static imbalance, \code{pr-journal} workload).
    The value of the parameter suggested by the original FSS algorithm is marked with a \textcolor{neonblue}{blue cross}, while the actual optimal solution targeted by our proposed method is marked with a \textcolor{neonblue}{blue star}.
    The error bands are the 95\% empirical bootstrap confidence intervals of the execution time mean.}\label{fig:param_space}
\end{figure*}

\subsection{Background}\label{background}
\noindent\textbf{Parallel loop scheduling.} Loops in scientific computing applications are easily parallelizable because of their embarrassingly-data-parallel nature.
A parallel loop scheduling algorithm attempts to map each task, or iteration, of a loop to CUs.
The most basic scheduling strategy called static scheduling (STATIC) equally divides the tasks (\( T_i \)) by the number of CUs in compile time.
Usually, a barrier is implied at the end of a loop, forcing all the CUs to wait until all tasks finish computing.
If imbalance is present across the tasks, some CUs may complete computation before other tasks, resulting in many CUs remaining idle.
Since execution time variance is abundant in practice because of control-flow divergence and inherent noise in modern computer systems~\cite{durand_impact_1996}, more advanced scheduling schemes are often required.

\vspace{0.05in}


\noindent\textbf{Dynamic loop scheduling.} Dynamic loop scheduling has been introduced to solve the inefficiency caused by execution time variance.
In dynamic scheduling schemes, each CU self-assigns a chunk of $K$ tasks in runtime by accessing a central task queue whenever it becomes idle.
The queue access causes a small runtime scheduling overhead, denoted by the constant $h$.
The case where $K=1$ is called self-scheduling (SS,~\cite{tang_processor_1986}).
For SS, we can achieve the minimum amount of load imbalance.
However, the amount of scheduling overhead grows proportionally to the number of tasks.
Even for small values of $h$, the total scheduling overhead can quickly become overwhelming.
The problem then boils down to finding the optimal tradeoff between load imbalance and scheduling overhead.
This problem has been mathematically formalized in~\cite{kruskal_allocating_1985, basthannah_provably_2000}, and a general review of the problem is provided in~\cite{yue_parallel_1995}.

\subsection{Factoring Self-Scheduling}\label{sched_algo}
Among many dynamic scheduling algorithms, we focus on the factoring self-scheduling algorithm (FSS,~\cite{hummel_factoring_1992}).
Instead of using a constant chunk size $K$, FSS uses a chunk size that decreases along the loop execution.
At the \(i\)th \textit{batch}, the size of the next $P$ chunks, $K_i$, is determined according to
\vspace{-0.02in}
\begin{align}
  R_0 &= N, \;\;\; R_{i+1} = R_i - P K_i, \;\;\; K_i = \frac{R_i}{x_i P} \label{eq:FSS} \\
  b_i &= \frac{P}{2 \sqrt{R_i}} \, \theta \label{eq:theta} \\ 
  x_0 &= 1 + b_0^2 + b_0 \sqrt{b_0^2 + 4} \\
  x_i &= 2 + b_i^2 + b_i \sqrt{b_i^2 + 4}.
\end{align}

\noindent where $R_i$ is the number of remaining tasks at the $i$th batch.
The parameter $\theta$ in~\eqref{eq:theta} is crucial to the overall performance of FSS.\
The analysis in~\cite{flynn_scheduling_1990} indicates that $\theta = \sigma / \mu$ results in the best performance where $\mu$ and $\sigma^2$ are the mean ($\mathbb{E}[T_i]$) and variance ($\mathbb{V}[T_i]$) of the tasks.
However, in Section~\eqref{motivation}, we show that this $\theta$ does not always perform well.
Instead, the essence of our work is a strategy to empirically determine good $\theta$ for each individual workloads by solving an optimization problem.

\vspace{0.05in}

\noindent\textbf{The FAC2 scheduling strategy.}
Since determining $\mu$ and $\sigma$ requires extensive profiling of the workload, the original authors of FSS suggest an unparameterized heuristic version~\cite{hummel_factoring_1992}.\
This version is often abbreviated as FAC2 in the literature and has been observed to outperform the original FSS~\cite{hagerup_allocating_1997, bast_scheduling_2000} despite being a heuristic modification.
Again, this observation supports the fact that the analytic solution \(\theta = \sigma / \mu\) is not always the best nor the only good solution.

\subsection{Motivation}\label{motivation}
\noindent\textbf{Limitations of workload-aware methods.}
The HSS and BinLPT strategies have significant drawbacks despite being able to fully incorporate the information about load imbalance.
First, both the HSS and BinLPT methods require an accurate workload-profile.
This is a significant limiting factor since many HPC workloads are comprised of homogeneous tasks where the imbalance is caused dynamically during runtime.
This means there is no static imbalance in the first place.
Also, even if a workload-profile is present, it imposes a runtime memory overhead of $O(N)$ for each loop.
For large-scale applications where the task count $N$ is huge, the memory overhead is a significant nuisance.

Moreover, both the HSS and BinLPT algorithms also have their own caveats.
The HSS algorithm has high scheduling overhead~\cite{penna_comprehensive_2019}.
In Section~\ref{eval}, we observe that HSS performs well only when high levels of imbalance, such as in the \code{pr-wiki} workload, are present.
On the other hand, BinLPT is highly sensitive to the accuracy of the workload-profile.
In practice, discrepancies between the actual workload and the workload-profile are inevitable.
We illustrate this fact using the \code{pr-journal} graph analytics workload in the upper plot of Fig.~\ref{fig:loadprof}.
We estimated the load of each task using the in-degree of the corresponding vertex in the graph.
The grey region is the estimated load of each task, while the \textcolor{neonred}{red region} is the measured load.
As shown in the figure, the estimated load does not accurately describe the actual load.
Likewise, the \textit{chunks} created by BinLPT using these estimates are equally inaccurate, as shown in the lower plot of Fig.~\ref{fig:loadprof}.
If the number of tasks is minimal, some level of discrepancy may be acceptable.
Indeed, the original analysis in~\cite{penna_comprehensive_2019} considers at most $N=3074$ tasks.
In practice, the number of tasks scales with data, leading to a very large $N$.

\vspace{0.05in}

\noindent\textbf{Effect of tuning the parameter of FSS.}
Similarly, classical scheduling algorithms such as FSS are not workload-robust~\cite{ciorba_openmp_2018}.
However, we reveal an interesting property by tuning the parameter (\(\theta\)) of FSS.
Fig.~\ref{fig:motiv1} and Fig.~\ref{fig:motiv2} illustrate the evaluation results of FSS using the \code{lavaMD} (a workload with low static imbalance) and \code{pr-journal} (a workload with high static imbalance) workloads with different values of \(\theta\), respectively.
The solution suggested in the original FSS algorithm (as discussed in Section~\ref{sched_algo}) is denoted by a \textcolor{neonblue}{blue cross}.
For the \code{lavaMD} workload (Fig.~\ref{fig:motiv1}), this solution is arguably close to the optimal value.
However, for the \code{pr-journal} workload (Fig.~\ref{fig:motiv2}), it leads to poor performance.
The original FSS strategy is thus not workload-robust since its performance varies greatly across workloads.

In contrast, by using an optimal value of \(\theta\) (\textcolor{neonblue}{blue star}), FSS outperforms all other algorithms as shown in the plots.
Even in Fig.~\ref{fig:motiv2} where HSS and BinLPT are equipped with an accurate workload-profile, FSS outperforms both methods.
This means that tuning the parameter of FSS on a per-workload basis can achieve robust performance.

\vspace{0.05in}

\noindent\textbf{Motivational remarks.} Workload-aware methods and classical dynamic scheduling methods tend to vary in applicability and performance.
Meanwhile, classic scheduling algorithms such as FSS achieve optimal performance when they are appropriately tuned to the target workload.
This performance potential of FSS points towards the possibility of creating a novel robust scheduling algorithm.


\section{Augmenting Factoring Self-Scheduling \\ with Bayesian Optimization}\label{method}
In this section, we describe BO FSS, a self-tuning variant of the FSS algorithm.
First, we provide an optimization perspective on the loop scheduling problem.
Next, we describe a solution to the optimization problem using BO.
Since solving our problem requires modeling of the execution time using surrogate models, we describe two ways to construct surrogate models. 

\subsection{Scheduling as an Optimization Problem}\label{problem}

The main idea of our proposed method is to design an optimal scheduling algorithm by finding its optimal configurations based on execution time measurements.
First, we define a set of scheduling algorithms $\mathcal{S} = \{ S_{\theta_1}, S_{\theta_2}, \ldots \}$ indentified by a tunable parameter $\theta$.
In our case, $\mathcal{S}$ is the set of configurations of the FSS algorithm with the parameter $\theta$ discussed in Section~\ref{sched_algo}.
Within this set of configurations, we choose the \textit{optimal configuration} that minimizes the mean of the total execution time contribution ($T_{total}$) of a parallel loop.
This problem is now of the form of an optimization problem denoted as

\begin{equation}
  \minimize_{\theta} \;\; \mathbb{E}[\, T_{total}(S_{\theta}) \,]. \label{eq:opt_problem}
\end{equation}

\noindent\textbf{Problem structure.} 
Now that the optimization problem is formulated, we are supposed to apply an optimization solver.
However, this optimization problem is ill-formed, prohibiting the use of any typical solver.
First, the objective function is noisy because of the inherent noise in computer systems.
Second, we do not have enough knowledge about the structure of $T$.
Different workloads interact differently with scheduling algorithms~\cite{ciorba_openmp_2018}.
It is thus difficult to obtain an analytic model of $T$ that is universally accurate.
Moreover, most conventional optimization algorithms require knowledge about the gradient $\nabla_{\theta}T$, which we do not have.
\begin{figure}[t]
\removelatexerror
\begin{algorithm}[H]
\small
\caption{Bayesian optimization}\label{alg:bo}
Initial dataset~~\(\mathcal{D}_0 = \{(\theta_0, \tau_0), \ldots, (\theta_N, \tau_N)\}\)
\For{\(t \in [1, T]\)}
{ 
  1. Fit surrogate model~~\(\mathcal{M}\) using \(\mathcal{D}_t\). \\
  2. Solve inner optimization problem. \( \theta_{t+1} = \argmax_{\theta} \; \alpha(\theta | \mathcal{M}, \mathcal{D}_t) \) \\
  3. Evaluate parameter.~~\( \tau_{t+1} \sim T_{\text{total}}(S_{\theta_{t+1}}) \) \\
  4. Update dataset.~~\(\mathcal{D}_{t+1} \leftarrow \mathcal{D}_{t+1} \cup (\theta, \tau)\)
}
\end{algorithm}
\end{figure}

\vspace{0.05in}

\begin{figure}[t]
  \centering
  \includegraphics[scale=0.21]{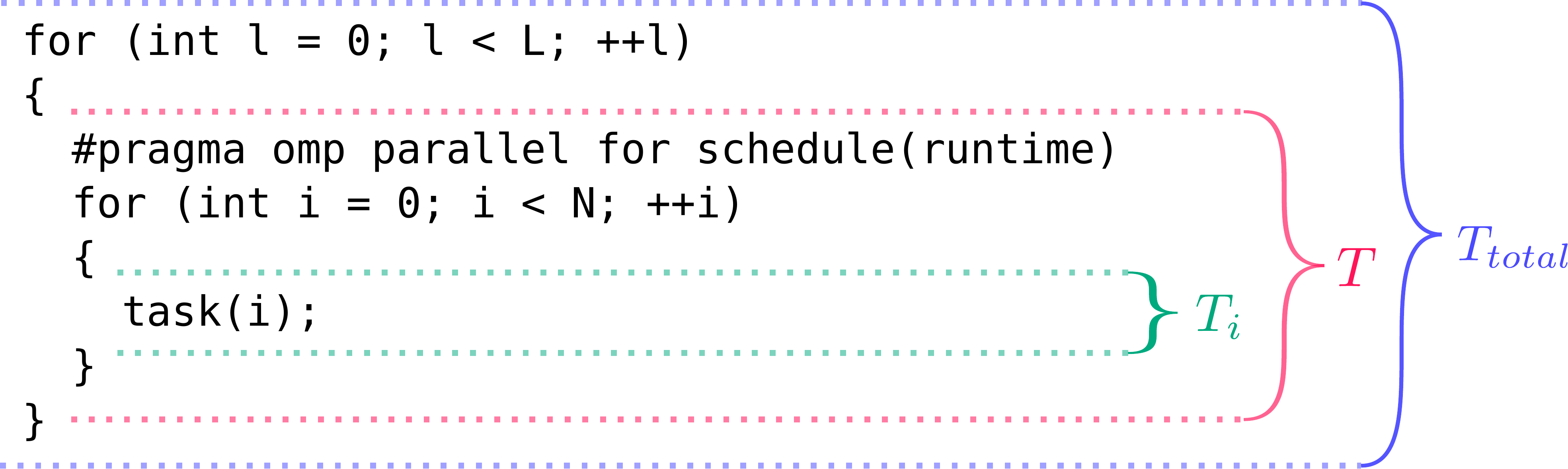}\label{fig:exec_time_diag}
  \caption{Visualization of our execution time models.
    The execution time of the parallel loop (\textcolor{neonred}{red bracket}) is denoted as \textcolor{neonred}{$T$}, while the execution time of the tasks in the parallel loop (\textcolor{neongreen}{green bracket}) is denoted as \textcolor{neongreen}{$T_i$}.
    The outer loop (\textcolor{neonblue}{blue bracket}) represents repeated execution ($L$ times) of the parallel loop within the application, where \textcolor{neonblue}{$T_{total}$} is the total execution time contribution of the loop.}
\end{figure}

\noindent\textbf{Solution using Bayesian optimization.}
For solving this problem, we leverage \textit{Bayesian Optimization} (BO).
We initially attempt to apply other gradient-free optimization methods such as stochastic approximation~\cite{spall_overview_1998}. However, the noise level in execution time is so extreme that most gradient-based methods fail to converge.
Conveniently, BO has recently been shown to be effective for solving such kind of optimization problems~\cite{alipourfard_cherrypick_2017, letham_constrained_2018, dalibard_boat_2017}.
Compared to other black-box optimization methods, BO requires less objective function evaluations and can handle the presence of noise well~\cite{alipourfard_cherrypick_2017}.

\vspace{0.05in}

\noindent\textbf{Description of Bayesian optimization.}
The overall flow of BO is shown in Algorithm~\ref{alg:bo}.
First, we build a surrogate model \(\mathcal{M}\) of $T_{\text{total}}$.
Let $(\theta, \tau)$ denote a \textit{data point} of an observation where \(\theta\) is a parameter value, and \(\tau\) is the resulting execution time measurement such that \(\tau \sim T_{\text{total}}\).
Based on a dataset of previous observations denoted as $\mathcal{D}_t = \{(\theta_1, \tau_1), \ldots, (\theta_t, \tau_t) \, \}$, a surrogate model provides a prediction of \(T_{\text{total}} (\theta)\) and the uncertainty of the prediction.
In our context, the prediction and uncertainty are given as the mean of the predictive distribution denoted as \(\mu(\theta \,|\, \mathcal{D}_t)\) and its variance denoted as \(\sigma^2(\theta \,|\, \mathcal{D}_t)\).

\begin{figure*}
  \centering
  \subfloat[Locality effect $\ell$-axis view]{
    \includegraphics[scale=0.4]{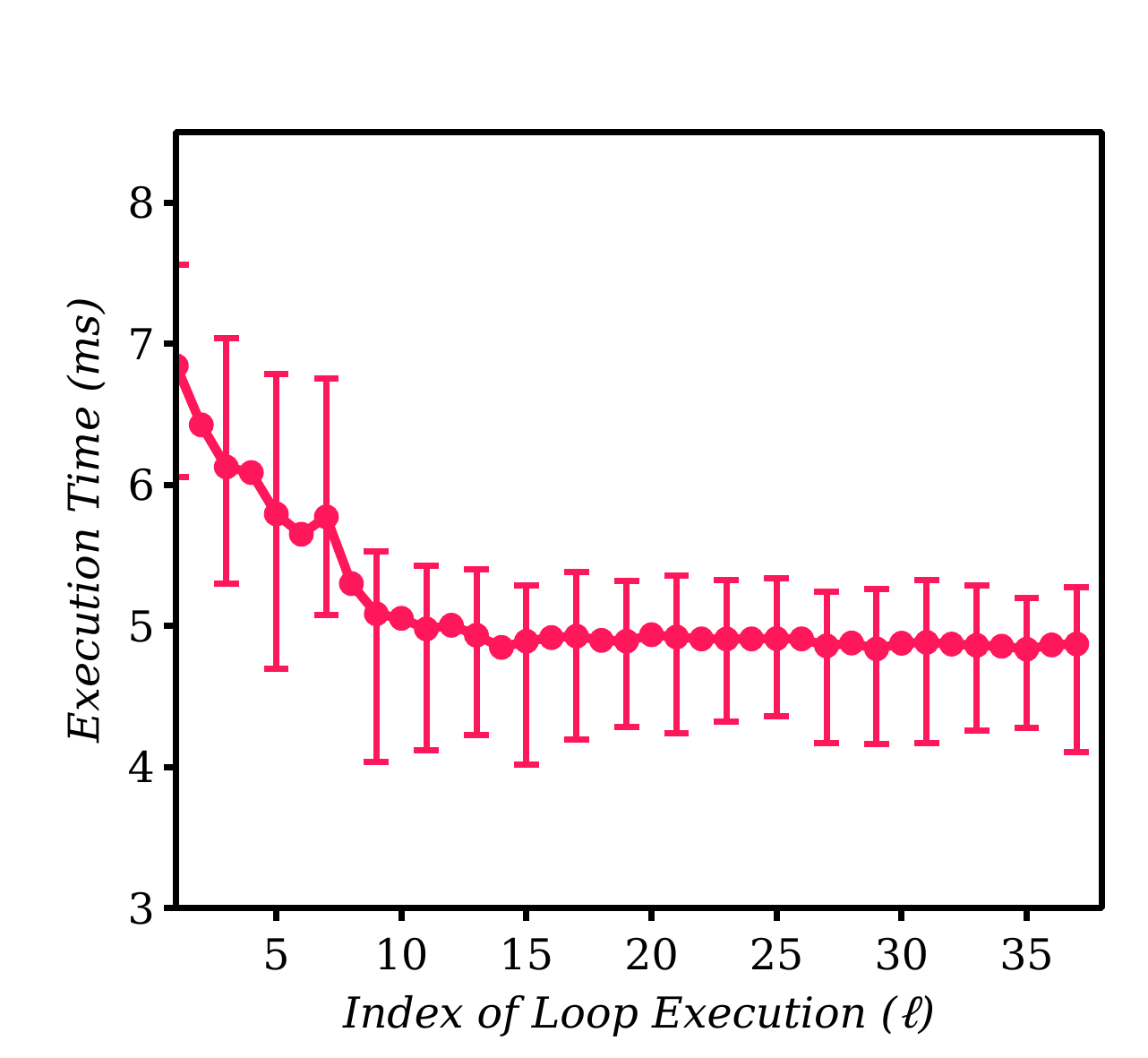}\label{fig:localityl}
  } 
  \subfloat[Locality effect $\theta$-axis view]{
    \includegraphics[scale=0.4]{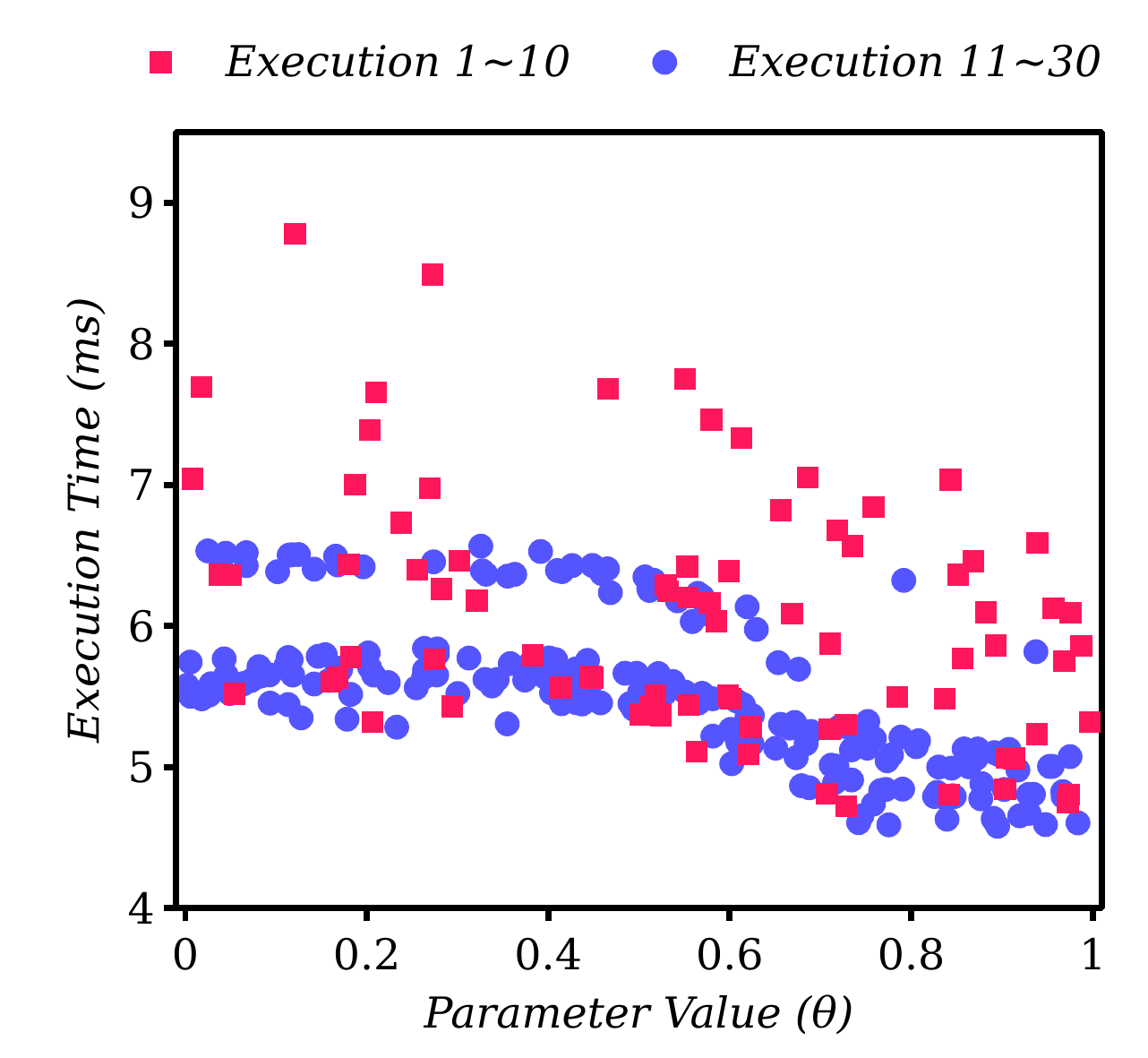}\label{fig:localitytheta}
  }
  \subfloat[Samples from the Exp.\ kernel]{
    \includegraphics[scale=0.4]{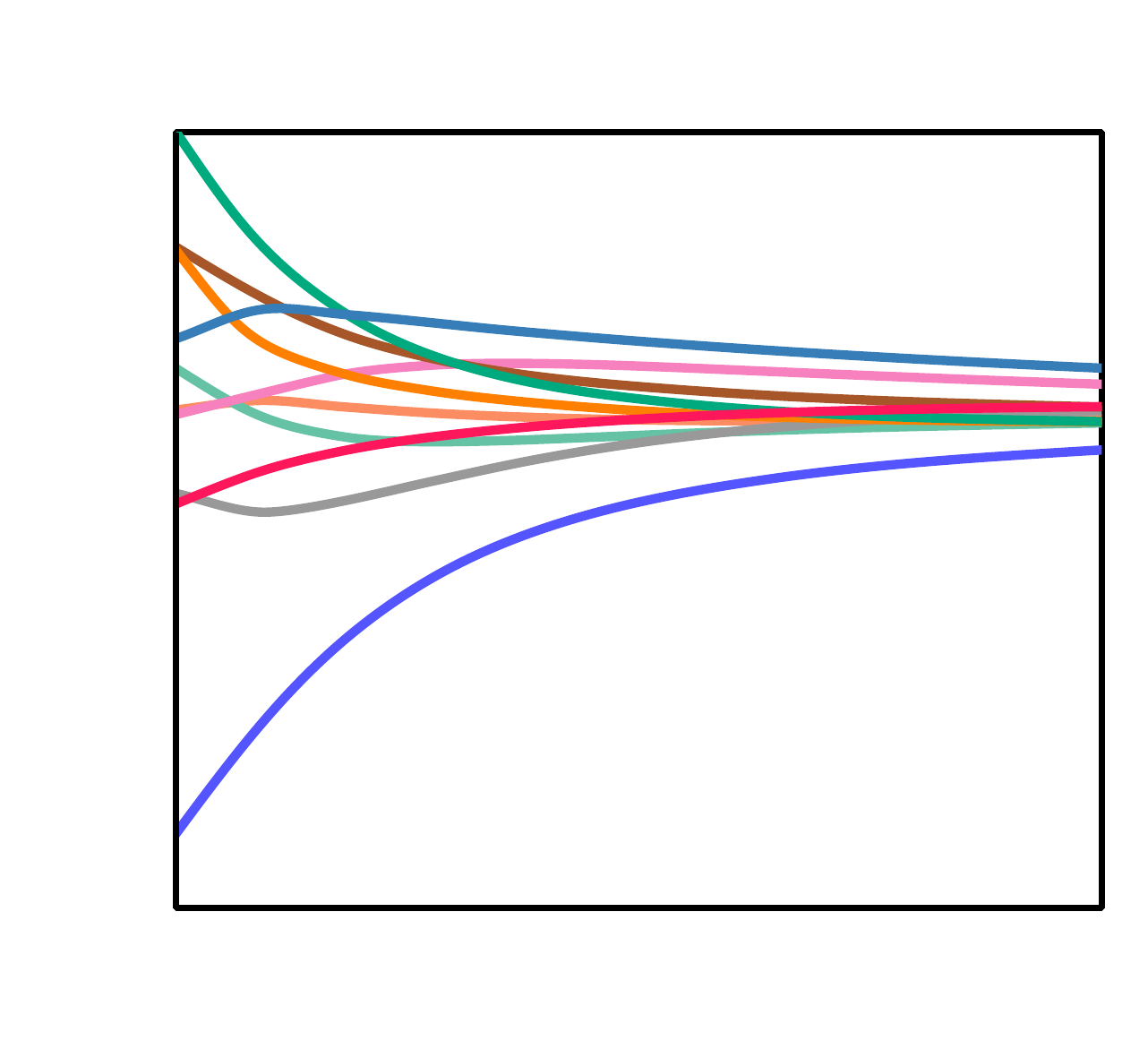}\label{fig:fsamples}
  }
  \caption{(a) (b) Visualization of the temporal locality effect on the execution time of the \code{kmeans} workload. (a) \(\ell\)-axis view. 
    The error bars are the 95\% empirical confidence intervals. (b) \(\theta\)-axis view. The \textcolor{red}{red squares} are measurements of earlier executions (\( \ell \leq 10 \)) while the \textcolor{neonblue}{blue circles} are measurements of later executions (\( \ell > 10 \)). (c) Randomly sampled functions from a GP prior with an exponentially decreasing function kernel.}\label{fig:models}
\end{figure*}

Using \(\mathcal{M}\), we now solve what is known as the \textit{inner optimization problem}.
In this step, we choose to \textit{exploit} our current knowledge about the optimal value or \textit{explore} entirely new values that we have not tried yet.
In the extremes, minimizing \(\mu(\theta \,|\, \mathcal{D}_t)\) gives us the optimal parameter \textit{given our current knowledge}, while minimizing \(\sigma^2(\theta \,|\, \mathcal{D}_t)\) gives us the parameter we are currently the most uncertain.
The optimal solution is given by a tradeoff of the two ends (often called the exploration-exploitation tradeoff), found by solving the optimization problem
%
\begin{align}
  \theta_{i+1} = \argmax_{\theta}{\, \alpha(\theta \,|\, \mathcal{M}, \mathcal{D}_t)}\label{eq:inner}
\end{align}
%
where the function \(\alpha\) is called the \textit{acquisition function}.
Based on the predictions and uncertainty estimates of \(\mathcal{M}\), \(\alpha\) returns our utility of trying out a specific value of \(\theta\).
Evidently, the quality of the prediction and uncertainty estimates of \(\mathcal{M}\) are crucial to the overall performance.
By maximizing \(\alpha\), we obtain the parameter value that has the highest utility, according to \(\alpha\).
In this work, we use the \textit{max-value entropy search} (MES,~\cite{wang_maxvalue_2017}) acquisition function.

After solving the inner optimization problem, we obtain the next value to try out, \(\theta_{t+1}\).
We can then try out this parameter and append the result (\(\theta_{t+1}, \tau_{t+1}\)) to the dataset.
For a comprehensive review of BO, please refer to~\cite{shahriari_taking_2016}.
We will later explain our \code{OpenMP} framework implementation of this overall procedure in Section~\ref{impl}.

\vspace{0.05in}

\subsection{Modeling Execution Time with Gaussian Processes}\label{total}
As previously stated, having a good surrogate model \(\mathcal{M}\) is essential.
Modeling the execution time of parallel programs has been a classic problem in the field of performance modeling.
It is known that parallel programs tend to follow a Gaussian distribution when the execution time variance is not very high~\cite{adve_influence_1993}.
This result follows from the \textit{central limit theorem} (CLT), which states that the influence of multiple \textit{i.i.d.} noise sources asymptotically form a Gaussian distribution.
Considering this, we model the total execution time contribution of a loop as
\begin{align}
  T_{total} &= \sum_{\ell = 1}^L \; T(S_\theta) + \epsilon \label{eq:model}
\end{align}

\noindent where $L$ is the total number of times a specific loop is executed within the application, indexed by $\ell$.
Following the conclusions of~\cite{adve_influence_1993}, we naturally assume that $\epsilon$ follows a Gaussian distribution.
Note that, at this point, we assume $T$ is independent of the index $\ell$.
For an illustration of the models used in our discussion, please see Fig.~\ref{fig:exec_time_diag}.

\vspace{0.05in}

\noindent\textbf{Gaussian Process formulation.}
From the dataset $\mathcal{D}_t$, we infer the model of the execution time $T_{total}(\theta)$ using \textit{Gaussian processes} (GPs).
A GP is a nonparametric Bayesian probabilistic machine learning model for nonlinear regression.
Unlike parametric models such as polynomial curve fitting and random forest, GPs automatically tune their complexity based on data~\cite{rasmussen_occam_2001}.
Also, more importantly, GPs can naturally incorporate the assumption of additive noise (such as \(\epsilon\) in~\eqref{eq:model}).
The prediction of a GP is given as a univariate Gaussian distribution fully described by its mean \( \mu(x | \mathcal{D}_t) \) and variance \( \sigma^2(x | \mathcal{D}_t) \).
These are computed in closed forms as
\begin{align}
  \mu(\theta | \mathcal{D}_t)      &= {\textbf{k}(\theta)}^T{(\mathbf{K} + \sigma_n^2I)}^{-1}\mathbf{y}\label{eq:mu} \\
  \sigma^2(\theta | \mathcal{D}_t) &= {k(\theta, \theta)} - {\textbf{k}(\theta)}^T{(\mathbf{K} + \sigma_\epsilon^2I)}^{-1} \, {\textbf{k}(\theta)} \label{eq:sigma}
\end{align}
\noindent where \(\mathbf{y} = [\tau_1, \tau_2, \ldots, \tau_t ] \), \(\mathbf{k}(\theta)\) is a vector valued function such that \({[\mathbf{k}(\theta)]}_i = k(\theta, \theta_i), \; \forall \theta_i \in \mathcal{D}_t\),  and \(\mathbf{K}\) is the Gram matrix such that \({[\, \mathbf{K} \, ]}_{i,j} = k(\theta_i, \theta_j), \; \forall \theta_i, \theta_j  \in \mathcal{D}_t\); \( k(x, y) \) denotes the \textit{covariance kernel} function which is a design choice.
We use the \textit{Matern 5/2} kernel which is computed as
\begin{align}
  k(x, x'; \, \sigma^2, \rho^2) &= \sigma^2 \, (1 + \sqrt{5}r + \frac{5}{3}r^2) \, \exp(-\sqrt{5}r) \label{eq:mat1}\\
           &\text{where} \;\; r = || x - x' ||_2 / \rho^2. \label{eq:mat2}
\end{align}
For a detailed introduction to GP regression, please refer to~\cite{rasmussen_gaussian_2006}.

\vspace{0.05in}

\begin{figure*}[h]
  \centering
  \includegraphics[scale=0.27]{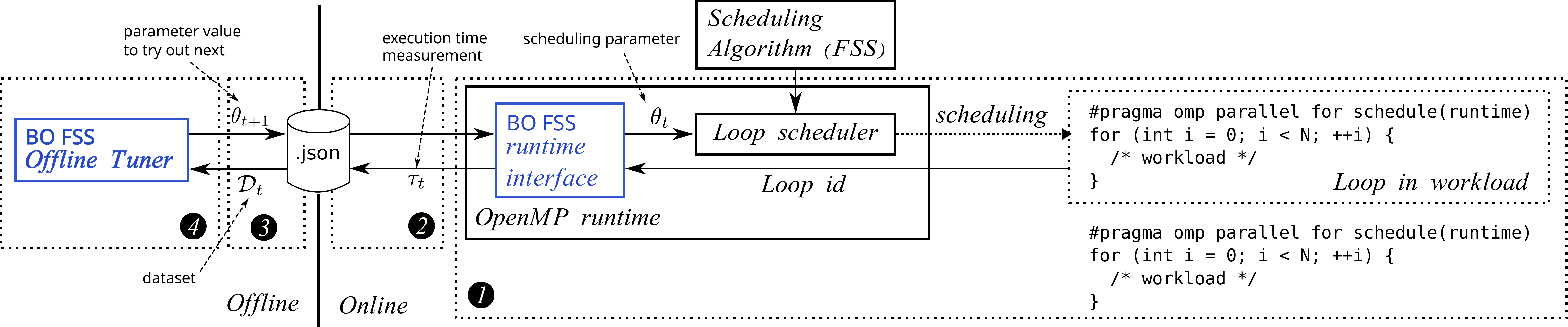}
  \caption{System overview of BO FSS.\
    \textit{Online} denotes the time we are actually executing the workload, while \textit{offline} denotes the time we are not executing the workload.
    For a detailed description, refer to the text in Section~\ref{impl}.
  }\label{fig:system}
\end{figure*}

\noindent\textbf{Non-Gaussian noise.}
Despite the remarks in~\cite{adve_influence_1993} saying that the noise in parallel programs mostly follow Gaussian distributions, we experienced cases where the execution time of individual parallel loops did not quite follow Gaussian distributions.
For example, occasional L2, L3 cache-misses results in large deviations, or \textit{outliers}, in execution time.
To correctly model these events, it is advisable to use heavy-tail distributions such as the Student-T.
More advanced methods for dealing with such outliers are described in~\cite{martinez-cantin_practical_2017} and~\cite{shah_bayesian_2013}.
However, to narrow the scope of our discussion, we stay within the Gaussian assumption.

\subsection{Modeling with Locality-Aware Gaussian Processes}\label{labo}
Until now, we only considered acquiring samples of $T_{total}$ by summing our measurements of $T$.
For the case where the parallel loop in question is executed more than once (that is, $L > 1$), we acquire $L$ observations of $T$ in a single run of the workload.
By exploiting our model's structure in~\eqref{eq:model}, it is possible to utilize all $L$ samples instead of aggregating them into a single one.
Since the Gaussian distribution is additive, we can decompose the distribution of $T_{total}$ such that
\begin{align}
  T_{total} &=  \sum_{\ell = 1}^L T(S_\theta, \ell) \\
          &\sim \sum_{\ell = 1}^L \mathcal{N}(\mathbb{E}[T(S_\theta, \ell)], \mathbb{V}[T(S_\theta, \ell)],) \\
          &=  \mathcal{N}(\, \sum_{\ell=1}^L \mathbb{E}[T(S_\theta, \ell)], \, \sum_{\ell=1}^L \mathbb{V}[T(S_\theta, \ell)] \,) \label{eq:exec_sum} \\
          &\approx  \mathcal{N}(\, \sum_{\ell=1}^L \mu(\theta, \ell \,|\, \mathcal{D}_t), \, \sum_{\ell=1}^L \sigma^2(\theta, \ell \,|\, \mathcal{D}_t) \,)\label{eq:labo}.
\end{align}
Note the dependence of \(T\) on the index of execution \(\ell\).
From~\eqref{eq:exec_sum}, we can retrieve $T_{total}$ from the mean ($\mathbb{E}[T(S_\theta, \ell)]$) and variance ($\mathbb{V}[T(S_\theta, \ell)]$) estimates of $T$, which are given by modeling $T$ using GPs as denoted in~\eqref{eq:labo}.

\vspace{0.05in}

\noindent\textbf{Temporal locality effect.}
However, this is not as simple as assuming that all $L$ measurements of $T$ are independent (ignoring the argument $\ell$ of $T$).
The execution time distribution of a loop changes dramatically within a single application run because of the temporal locality effect.
This is shown in Fig.~\ref{fig:models} using measurements of a loop in the \code{kmeans} benchmark.
In Fig.~\ref{fig:localityl}, it is clear that earlier executions of the loop (\( \ell \leq 10 \)) are much longer than the later executions (\( \ell > 10 \)).
Also, different moments of executions are effected differently by \(\theta\), as shown in Fig.~\ref{fig:localitytheta}.
It is thus necessary to accurately model the effect of \(\ell\) to better distinguish the effect of \(\theta\).

\vspace{0.05in}

\noindent\textbf{Exponentially decreasing function kernel.}
To model the temporal locality effect, we expand our GP model to include the index of execution $\ell$.
Now, the model is a 2-dimensional GP receiving $\ell$ and $\theta$.
Within the workloads we consider, the temporal locality effect is shown an exponentially decreasing tendency.
We thus assume that the locality effect can be represented with \textit{exponentially decreasing functions} (Exp.) of the form of $e^{-\lambda \ell}$.
The kernel for these functions has been introduced in~\cite{swersky_freezethaw_2014} for modeling the learning curves of machine learning algorithms.
The exponentially decreasing function kernel is computed as
\begin{align}
  k(\ell, \ell') = \frac{\beta^\alpha}{{(\ell + \ell' + \beta)}^\alpha}.\label{eq:expkern}
\end{align}
Random functions sampled from the space induced by the Exp. kernel are shown in Fig.~\ref{fig:fsamples}.
Notice the similarity of the sampled functions and the visualized locality effect in Fig.~\ref{fig:localityl}.
Modeling more complex locality effects such as periodicity can be achieved by combining more different kernels.
An automatic procedure for doing this is described in~\cite{duvenaud_structure_2013}.

\vspace{0.05in}

\noindent\textbf{Kernel of locality-aware GPs.}
Since the sum of covariance kernels is also a valid covariance kernel~\cite{rasmussen_gaussian_2006}, we define our 2-dimensional kernel as
\begin{align}
  &k(x, x') = k_{\text{Matern}} (\theta, \theta') + k_{\text{Exp}} (\ell, \ell') \label{eq:sumker} \\
  &\textit{where}\;\; x = [\,\theta,\, \ell \,], \; x' = [\,\theta',\, \ell' \,].
\end{align}
Intuitively, this definition implies that we assume the effect of scheduling (resulting from $\theta$) and locality (resulting from $\ell$) to be superimposed (additive).

\vspace{0.05in}
\noindent\textbf{Reducing computational cost.}
The computational complexity of computing a GP is in $O(T^3)$ where $T$ represents the total number of BO iterations.
The locality aware construction uses all the independent loop executions resulting in computational complexity in $O({(LT)}^3)$.
To reduce the computational cost, we subsample data along the axis of $\ell$ by using every $k$th measurement of the loop, such that $\ell \in \{1,\, k + 1,\, 2k + 1,\, \ldots,\, L\}$.
As a result, the computational complexity is reduced by a constant factor such that $O\left({\left(\frac{L}{k}T\right)}^3\right)$.
In all of our experiments, we use a large value of \(k\) so that \(L/k = 4\).

\subsection{Treatment of Gaussian Process Hyperparameters}\label{hyper}
GPs have multiple hyperparameters that need to be predetermined.
The suitability of these hyperparameters is directly related to the optimization performance of BO~\cite{henrandez-lobato_predictive_2014}.
Unfortunately, whether a set of hyperparameters is appropriate depends on the characteristics of the workload. 
Since real-life workloads are very diverse, it is essential to automatically handle these parameters.

The Matern 5/2 kernel has two hyperparameters $\rho$ and $\sigma$, while the exponentially decreasing function kernel has two hyperparameters $\alpha$ and $\beta$.
GPs also have hyperparameters themselves, the function mean $\mu$ and the noise variance $\sigma^2_\epsilon$.
We denote the hyperparameters using the concatenation $\phi = [\, \mu, \sigma_\epsilon, \sigma, \rho, \ldots \,]$.

Since the marginal likelihood $p(\mathcal{D}_t|\phi)$ is available in a closed form~\cite{rasmussen_gaussian_2006}, we can infer the hyperparameters using maximum likelihood estimation type-II or marginalization by integrating out the hyperparameters.
Marginalization has empirically shown to give better optimization performance in the context of BO~\cite{snoek_practical_2012, henrandez-lobato_predictive_2014}.
It is performed by approximating the integral
\begin{align}
  \alpha(x \,|\, \mathcal{M}, \mathcal{D}_t) &= \int \alpha(x \,|\, \mathcal{M}, \phi, \mathcal{D}_t) p(\phi | \mathcal{D}_t) d\theta \label{eq:bayes} \\
                           &\approx \frac{1}{N}\sum_{\phi_i \sim p(\phi | \mathcal{D}_t)} \alpha(x \,|\, \mathcal{M}, \phi_i, \mathcal{D}_t), \label{eq:mcmc}
\end{align}
using samples from the posterior $\phi_i$ where \(N\) is the number of samples.
For sampling from the posterior, we use the no-u-turn sampler (NUTS,~\cite{JMLR:v15:hoffman14a}).


\begin{table*}[t]
  \centering
\begin{threeparttable}
  \caption{Benchmark Workloads}\label{table:real}
\begin{tabular}{lllrlc}
  \toprule
 \textbf{Suite}  & \textbf{Workload Profile} & \textbf{Characterization} & \textbf{\# Tasks ($N$)} & \textbf{Application Domain} & \textbf{Benchmark Suite}  \\
  \midrule
 \code{lavaMD}  & Uniformative \tnote{1} & N-Body                 & 8000                  & Molecular Dynamics & Rodinia 3.1 \\
 \code{stream.} & No                     & Dense Linear Algebra   & 65536                 & Data Mining        & Rodinia 3.1\\
 \code{kmeans}  & Uniformative \tnote{2} & Dense Linear Algebra   & 494020                & Data Mining        & Rodinia 3.1\\
 \code{srad v1} & Uniformative \tnote{1} & Structured Grid        & 229916                & Image Processing   & Rodinia 3.1\\
 \code{nn}      & Uniformative \tnote{1} & Dense Linear Algebra   & 8192                  & Data Mining        & Rodinia 3.1\\
 \code{cc-}*    & Yes                    & Sparse Linear Algebra  & N/A \tnote{3}         & Graph Analytics    & GAP    \\
 \code{pr-}*    & Yes                    & Sparse Linear Algebra  & N/A \tnote{3}         & Graph Analytics    & GAP  \\
 \bottomrule
\end{tabular}
\begin{tablenotes}
\item[1] Uniformly partitioned workload.
\item[2] Imbalance present only in domain boundaries.
\item[3] Input data dependent; number of vertices of the input graph.
\end{tablenotes}
\end{threeparttable}\\
\end{table*}

\section{System Implementation}\label{impl}
We now describe our implementation of BO FSS.
Our implementation is based on the GCC implementation of the \code{OpenMP} 4.5 framework~\cite{dagum_openmp_1998}, which is illustrated in Fig.~\ref{fig:system}.
The overall workflow is as follows:
\begin{enumerate}[label=\protect\circleddark{black}{\arabic*}]
  \addtocounter{enumi}{-1}
  \item First, we randomly generate initial scheduling parameters \(\theta_0, \ldots, \theta_{N_0} \) using a Sobol quasi-random sequence~\cite{sobol_distribution_1967}.
  \item During execution, for each loop in the workload, we schedule the parallel loop using the parameter \( \theta_t \).
  We measure the resulting execution time of the loop and acquire a measurement \(\tau_t\).
  \item Once we finish executing the workload, store the pair \((\theta_t, \tau_t)\) in disk in a \code{JSON} format.
  \item Then, we run the offline tuner, which loads the dataset \(\mathcal{D}_t\) from disk.
  \item Using \(\mathcal{D}_t\), we solve the inner optimization problem in~\eqref{eq:inner}, obtaining the next scheduling configuration \(\theta_{t+1}\).
  \item At the subsequent execution of the workload, \( t \leftarrow t + 1\), and go back to \circleddark{black}{1}.
\end{enumerate}
Note that \textit{offline} means the time we finish executing the workload, while \textit{online} means the time we are executing the workload (runtime).

\vspace{0.02in}

\noindent\textbf{Implementation of the offline tuner.}
We implemented the offline tuner as a separate program written in Julia~\cite{bezanson_julia_2017}, which is invoked by the user.
When invoked, the tuner solves the inner optimization problem, and stores the results in disk.
For solving the inner optimization problem, we use the \textit{DIRECT} algorithm~\cite{jones_lipschitzian_1993} implemented in the NLopt library\cite{johnson_nlopt_2011}.
For marginalizing the GP hyperparameters, we use the \code{AdvancedHMC.jl} implementation of NUTS\cite{ge2018t}.

\vspace{0.02in}

\noindent\textbf{Search space reparameterization.}
BO requires the domain of the parameter to be bounded.
However, in the case of FSS, \( \theta \) is not necessarily bounded.
As a compromise, we reparameterized \( \theta \) into a fixed domain such that

\vspace{-0.1in}
\begin{align}
  &\minimize_{x} \;\; \mathbb{E}[\, T_{total}(S_{\theta(x)}) \,] \\
  &\;\; \text{\textit{where}} \;\; \theta(x) = 2^{19 \, x - 10}, \; 0 < x < 1. 
\end{align}
This also effectively converts the search space to be in a logarithmic scale.
The reparameterized domain was determined by empirically investigating feasible values of \( \theta \).

\vspace{0.05in}

\noindent\textbf{User interface.} BO FSS can be selected by setting the \code{OMP\_SCHEDULE} environment variable, or by the \code{OpenMP} runtime API as in Listing~\ref{omp_sched_select}.

\vspace{-0.2in}
\begin{lstfloat}
\begin{lstlisting}[
    caption=Selecting a scheduling algorithm,
    label=omp_sched_select,
    escapechar=@,
    style=CABI ]
omp_set_schedule(BO_FSS); // selects BO FSS
\end{lstlisting}
\end{lstfloat}
\vspace{-0.2in}

\noindent\textbf{Modification of the OpenMP ABI.}
As previously described, our system optimizes each loop in the workload independently.
Naturally, our system requires the identification of the individual loops within the \code{OpenMP} runtime.
However, we encountered a major issue: the current \code{OpenMP} ABI does not provide a way for such identification.
Consequently, we had to modify the GCC 8.2~\cite{gcc_gcc_2018} compiler's \code{OpenMP} code generation and the \code{OpenMP} ABI.\
The modified GCC \code{OpenMP} ABI is shown in Listing~\ref{omp_abi}.
During compilation, a unique token for each loop is generated and inserted at the end of the \code{OpenMP} procedure calls.
Using this token, we store and manage the state of each loop.
Measuring the loop execution time is done by starting the system clock in \code{OpenMP} runtime entries such as \code{GOMP\_parallel\_runtime\_start}, and stopping in exits such as \code{GOMP\_parallel\_end}.

\vspace{-0.2in}
\begin{lstfloat}
\noindent\begin{lstlisting}[
  caption={Modified GCC OpenMP ABI},
  label=omp_abi,
  escapechar=@,
  style=CABI ]
void GOMP_parallel_loop_runtime(void (*fn) (void *), void *data, unsigned num_threads, long start, long end, long incr, unsigned flags, size_t @\color{neonblue}{\textbf{loop\_id}}@)
void GOMP_parallel_runtime_start(long start, long end, long incr, long *istart, long *iend, size_t @\color{neonblue}{\textbf{loop\_id}}@)
void GOMP_parallel_end(size_t @\color{neonblue}{\textbf{loop\_id}}@)
\end{lstlisting}
\end{lstfloat}
\vspace{-0.4in}



\begin{table*}[t]
  \vspace{-0.1in}
  \centering
  \caption{Minimax Regret of Scheduling Algorithms}\label{table:regret}
  \vspace{-0.1in}
\begin{tabular}{lrrrrrrrrrr}
\multicolumn{11}{p{0.8\linewidth}}{ The values in the table are the percentage slowdown relative to the best performing algorithm.
  They can be interpreted as the opportunity cost of using each algorithm. 
  For more details, refer to the text in Section~\ref{setup}.
  \vspace{0.1in}
} \\
  
\toprule
\multirowcell{2}{\textbf{Workload}} & \multicolumn{1}{c}{\textbf{\textit{Ours}}}   & \multicolumn{1}{c}{\textbf{Static}} & \multicolumn{2}{c}{\textbf{Workload-Aware}}                   & \multicolumn{6}{c}{\textbf{Dynamic}} \\
\cmidrule(lr){2-2} \cmidrule(lr){3-3} \cmidrule(lr){4-5} \cmidrule(lr){6-11}
& \multicolumn{1}{c}{\textbf{BO FSS}} & \multicolumn{1}{c}{\textbf{STATIC}} & \multicolumn{1}{c}{\textbf{HSS}} & \multicolumn{1}{c}{\textbf{BinLPT}} & \multicolumn{1}{c}{\textbf{GUIDED}} & \multicolumn{1}{c}{\textbf{FSS}} & \multicolumn{1}{c}{\textbf{CSS}} & \multicolumn{1}{c}{\textbf{FAC2}} & \multicolumn{1}{c}{\textbf{TRAP1}} & \multicolumn{1}{c}{\textbf{TAPER3}} \\
\midrule
\code{lavaMD}                               & \textbf{0.00}                        & 17.55                      &                        n/a &  n/a                          & 7.25                       & 3.00                    & 0.36                   & 0.25                     & 10.33                     & 42.64                      \\
\code{stream}.                              & \textbf{0.00}                        & 10.79                     &                         n/a&                            n/a& 2.39                       & 10.36                   & 1.25                    & 0.68                     & 2.00                      & 2.45                       \\
\code{kmeans}                               & \textbf{0.00}                       & 23.02                      &n/a                         &                           n/a & 8.01                       & 17.62                   & 1.50                    & 1.17                     & 2.30                      & 6.41                       \\
\code{srad v1}                              & 22.34                      & 10.92                      &                         n/a& n/a                           & 16.75                      & 11.74                   & 26.03                   & \textbf{0.00}                     & 16.43                     & 17.61                      \\
\code{nn}                                   & 4.76                       & 5.06                       &                         n/a&                           n/a & \textbf{0.00}                       & 0.55                    & 7.00                    & 6.06                     & 4.39                      & 5.14                       \\
\code{cc-journal}                           & \textbf{0.00}                        & 2.88                       & 66.98                   & 196.63                     & 11.94                      & 2.47                    & 2.98                    & 6.15                     & 3.65                      & 0.66                       \\
\code{cc-wiki}                              & \textbf{0.00}                        & 6.94                       & 58.57                   & 154.31                     & 10.37                      & 2.77                    & 6.58                    & 5.29                     & 7.88                      & 5.27                       \\
\code{cc-road}                              & \textbf{0.00}                        & 8.57                       & 81.88                   & 251.71                     & 7.19                       & 1.37                    & 1.55                    & 1.23                     & 1.97                      & 1.71                       \\
\code{cc-skitter}                           & 5.28                       & 2.28                       & 61.69                   & 129.08                     & 3.57                       & 1.03                    & 1.05                    & 1.06                     & 0.73                      & \textbf{0.00}                       \\
\code{pr-journal}                           & \textbf{0.00}                        & 29.66                      & 5.52                    & 66.89                      & 42.93                      & 29.01                   & 29.07                   & 29.17                    & 29.33                     & 28.81                      \\
\code{pr-wiki}                              & 15.30                      & 45.20                      & \textbf{0.00}                     & 42.26                      & 85.34                      & 46.99                   & 47.28                   & 46.82                    & 46.53                     & 46.87                      \\
\code{pr-road}                              & \textbf{0.00}                        & 0.32                       & 41.65                   & 138.32                     & 6.60                       & 0.41                    & 0.42                    & 0.42                     & 0.40                      & 0.41                       \\
\code{pr-skitter}                           & \textbf{0.00}                        & 11.51                      & 23.21                   & 68.91                      & 29.97                      & 11.66                   & 11.21                   & 11.34                    & 12.06                     & 11.26                      \\
\midrule
\multicolumn{1}{c}{ \( \mathcal{R}(S) \)}  & \textbf{22.34}                      & 45.20                      & 81.88                   & 251.71                     & 85.34                      & 46.99                   & 47.28                   & 46.83                    & 46.53                     & 46.87  \\
\multicolumn{1}{c}{ \( \mathcal{R}_{90}(S) \)}  & \textbf{13.30} & 28.33 & 71.75 & 213.15 & 40.34 & 26.73 & 28.46 & 25.60 & 26.75 & 39.87  \\
\bottomrule
\end{tabular}
\end{table*}

\section{Evaluation}\label{eval}
In this section, 
we first describe the overall setup of our experiments.
Then, we compare the robustness of BO FSS against other scheduling algorithms.
After that, we evaluate the performance of our BO augmentation scheme.
Lastly, we directly compare the execution time.

\subsection{Experimental Setup}\label{setup}
\noindent\textbf{System setup.}
All experiments are conducted on a single shared-memory computer with an AMD Ryzen Threadripper 1950X 3.4GHz CPU which has 16 cores and 32 threads with simultaneous multithreading enabled.
It also has 1.5MB of L1 cache, 8MB of L2 cache and 32MB of last level cache.
We use the Linux 5.4.36-lts kernel with two 16GB DDR4 RAM (32GB total).
Frequency scaling is disabled with the \code{cpupower frequency-set performance} setting.
We use the GCC 8.3 compiler with the \code{-O3}, \code{-march=native} optimization flags enabled in all of our benchmarks.

\vspace{0.05in}
\noindent\textbf{BO FSS setup.}
We run BO FSS for 20 iterations starting from 4 random initial points.
All results use the best parameter found after the aforementioned number of iterations.

\vspace{0.05in}

\noindent\textbf{Baseline scheduling algorithms.}
We compare BO FSS against the FSS~\cite{hummel_factoring_1992}, CSS~\cite{kruskal_allocating_1985}, TSS~\cite{tzen_trapezoid_1993}, GUIDED~\cite{polychronopoulos_guided_1987}, TAPER~\cite{lucco_dynamic_1992}, BinLPT~\cite{penna_comprehensive_2019}, HSS~\cite{kejariwal_historyaware_2006} algorithms.
%
%
We use the implementation of BinLPT and HSS provided by the authors of BinLPT\footnote{Retrieved from \url{https://github.com/lapesd/libgomp}}.
For the FSS and CSS algorithms, we estimate the statistics of each workloads ($\mu$, $\sigma$) beforehand from 64 executions.
The scheduling overhead parameter $h$ is estimated using the method described in~\cite{bull_measuring_1999}.
We use the default STATIC and GUIDED implementations of the \code{OpenMP} 4.5 framework using the \code{static} and \code{guided} scheduling flags.
For the TSS and TAPER schedules, we follow the heuristic versions suggested in their original works, denoted as TRAP1 and TAPER3, respectively.

\vspace{0.05in}
\noindent\textbf{Benchmark workloads.}
The workloads considered in our experiments are summarized in Table~\ref{table:real}.
We select workloads from the Rodinia 3.1 benchmark suite~\cite{che_characterization_2010} (\code{lavamd}, \code{streamcluster}, \code{kmeans}, \code{srad v1}) where the STATIC scheduling method performs worse than other dynamic scheduling methods. 
We also include workloads from the GAP benchmark suite~\cite{beamer_gap_2017} (\code{cc}, \code{pr}) where the load is predictable from the input graph.

\vspace{0.05in}

\noindent\textbf{Workload-profile availability.}
We characterize the workload-profile availability of each workload in the \textit{Workload-Profile} column in Table~\ref{table:real}.
For workloads with homogeneous tasks (\code{lavaMD}, \code{stream.}, \code{srad v1}, \code{nn}), static imbalance does not exist.
Most of the imbalance is caused during runtime, deeming a workload-profile uniformative.
On the other hand, the static imbalance of the \code{kmeans} workload is revealed during execution, not before execution.
We thus consider the workload-profile to be effectively unavailable.

\begin{figure*}
  \vspace{-0.1in}
  \centering
  \includegraphics[scale=0.45]{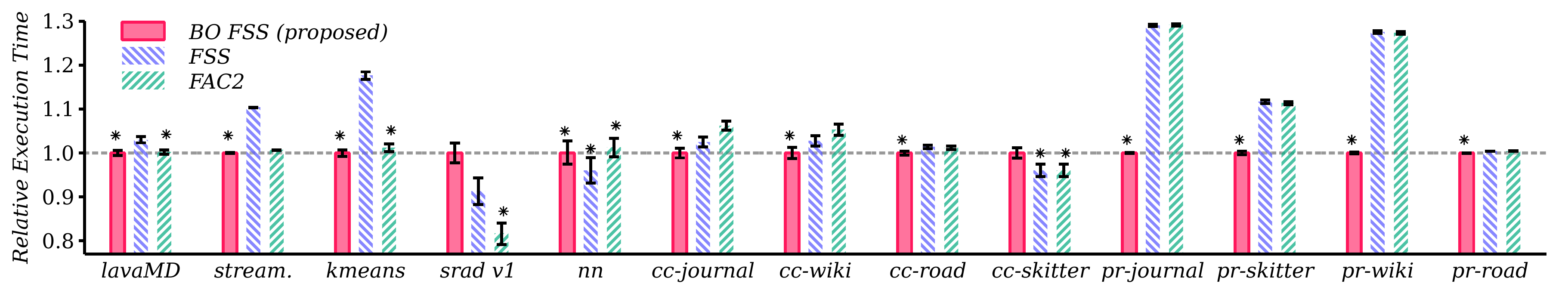}
  \vspace{-0.1in}
  \caption{Execution time comparison of BO FSS, FSS and FAC2. We estimate the mean execution time from 256 executions. The error bars show the 95\% bootstrap confidence intervals. The results are normalized by the mean execution time of BO FSS. The methods with the lowest execution time are marked with a star (*). Methods \textit{not} significantly different with the best performing method are also marked with a star (Wilcoxon signed rank test, 1\% null-hypothesis rejection threshold).}\label{fig:fssfamily}
\end{figure*}

\begin{table}[ht]
  \footnotesize
  \begin{threeparttable}
  \caption{Input Graph Datasets}\label{table:graphs}
\begin{tabular}{lrrccc}
\toprule
\multirowcell{2}{\textbf{Dataset}} & \multirowcell{2}{\textbf{$\mathbf{| \mathcal{V} | }$}} & \multirowcell{2}{\textbf{$\mathbf{| \mathcal{E} |}$}} & 
\multicolumn{3}{c}{$\mathbf{\text{\textbf{deg}}^-(v)}$\tnote{1}, $\mathbf{\text{\textbf{deg}}^+(v)}$\tnote{2}} \\
\cmidrule{4-6}
& & & {\textbf{mean}} & {\textbf{std}} & {\textbf{max}} \\ 
\midrule
journal~\cite{backstrom_group_2006}        & 4.0M  & 69.36M & 17, 17 & 43, 43 & 15k, 15k \\
wiki~\cite{gleich_wikipedia20070206_2007} & 3.57M & 45.01M & 13, 13 & 33, 250 & 7k, 187k\\
road~\cite{demetrescu_9th_2006} & 24.95M & 57.71M & 2, 2 & 1, 1 & 9, 9 \\
skitter~\cite{leskovec_graphs_2005} & 1.70M & 22.19M & 13, 13 & 137, 137 & 35k, 35k \\
\bottomrule
\end{tabular}
\begin{tablenotes}
\item[1] In-degree of each vertex.
\item[2] Out-degree of each vertex.
\end{tablenotes}
  \end{threeparttable}
\end{table}

\vspace{0.05in}

\noindent\textbf{Input graph datasets.}
We organize the graph datasets used for the workloads from the GAP benchmark suite in Table~\ref{table:graphs}, acquired from~\cite{davis_university_2011}.
$|\mathcal{V}|$ and $|\mathcal{E}|$ are the vertices and edges in each graph, respectively.
The load of each task $T_i$ in the \code{cc} and \code{pr} workloads is proportional to the in-degree and out-degree of each vertex~\cite{bak_optimized_2019}.
We use this degree information for forming the workload-profiles.
Among the datasets considered, \code{wiki} has the most extreme imbalance while \code{road} has the least imbalance~\cite{bak_optimized_2019}.

\vspace{0.05in}

\noindent\textbf{Workload-robustness measure.}
To quantify the notion of workload-robustness, we use the \textit{minimax regret} measure~\cite{doi:10.1080/01621459.1951.10500768}.
The minimax regret quantifies robustness by calculating the opportunity cost of using an algorithm, computed as

\begin{align}
  &\mathcal{R}(S, w) = \frac{C(S, w) - \min_{S \in \mathcal{S}} C(S, w)}{\min_{S \in \mathcal{S}} C(S, w)} \times 100 \label{eq:regret} \\
  &\mathcal{R}(S)    =  \max_{w \in \mathcal{W}} \mathcal{R}(S, w)\label{eq:final_regret}
\end{align}

\noindent where \(C(S, w)\) is the cost of the scheduling algorithm \(S\) on the workload \(w\), and \(\mathcal{W}\) is our set of workloads.
We choose \(C(S, w)\) to be the execution time.
In this case, \(\mathcal{R}(\mathcal{S}, w)\) can be interpreted as the slowdown relative to the best performing algorithm in percentages.
Also, \(\mathcal{R}(\mathcal{S})\) is the worst case regret of using \(\mathcal{S}\) on the set of workloads \(\mathcal{W}\).
Note that among different robustness measures, the minimax regret is very pessimistic~\cite{mcphail_robustness_2018}, emphasizing worst-case performance.
For this reason, we additionally consider the 90th percentile of the minimax regret denoted as \(\mathcal{R}_{90}(S)\).

\subsection{Evaluation of Workload-Robustness}\label{robust}
Table~\ref{table:regret} compares the minimax regrets of different scheduling algorithms with that of BO FSS.
Each entry in the table is the regret subject to the workload and scheduling algorithm, \(\mathcal{R}(S, w)\).
The final rows are the minimax regret~\(\mathcal{R}(S)\) and the 90th percentile minimax regret~\(\mathcal{R}_{90}(S)\) subject to the scheduling algorithm.
BO FSS achieves the lowest regret both in terms of minimax regret (\(22\%\) points) and 90th percentile minimax regret (\(13\%\) points).
In contrast, both static and dynamic scheduling methods achieve similar level of regret.
This observation is on track with the previous findings~\cite{ciorba_openmp_2018}; none of the classic scheduling methods dominate each other.

It is worth to note that we selected workloads in which STATIC performs poorly.
Our robustness analysis thus only holds for comparing dynamic and workload-aware scheduling methods.

\vspace{0.05in}

\noindent\textbf{Remarks.}
The results for workload-robustness using the minimax regret metric show that BO FSS achieves significantly lower levels of regret compared to other scheduling methods. As a result, BO FSS performs consistently well.
Even when BO FSS does not perform the best, its performance is within an acceptable range.

\subsection{Evaluation of Bayesian Optimization Augmentation}\label{perf_aug}
A fundamental part of the proposed method is that BO FSS improves the performance of FSS by tuning its internal parameter.
In this section, we show how much BO augmentation improves the performance of FSS and its heuristic variant FAC2.
We run BO FSS, FSS, and FAC2 on workloads with both high and low static imbalances.
The results are shown in Fig.~\ref{fig:fssfamily}.
Overall, we can see that BO FSS consistently outperforms FSS and FAC2 with the exception of \code{srad v1} and \code{cc-skitter}.
On workloads with high imbalance such as \code{pr-journal} and \code{pr-wiki}, the execution time improvements are as high as 30\%.
%
%
\begin{figure}[t]
  \vspace{-0.2in}
  \centering
  \includegraphics[scale=0.40]{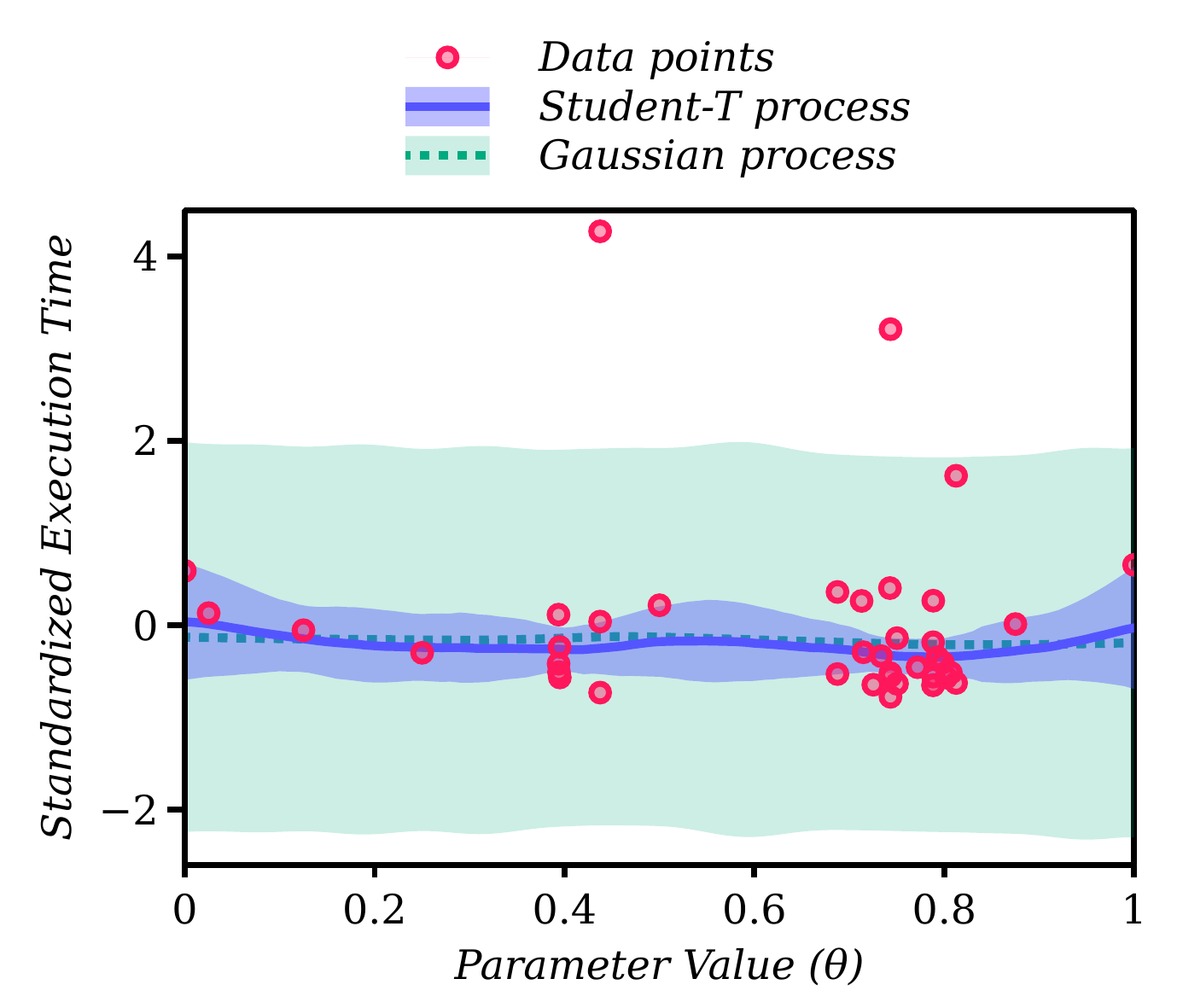}
  \vspace{-0.1in}
  \caption{Parameter space and surrogate model fit on the \code{srad v1} workload.
    The colored regions are the 95\% predictive credible intervals of a GP (\textcolor{neongreen}{green region}) and a Student-T process (\textcolor{neonblue}{blue region}).
    The \textcolor{red}{red circles} are the data points used to fit both surrogate models.
  }\label{fig:outlier}
\end{figure}

\begin{figure}[t]
  \vspace{-0.2in}
  \centering
  \includegraphics[scale=0.40]{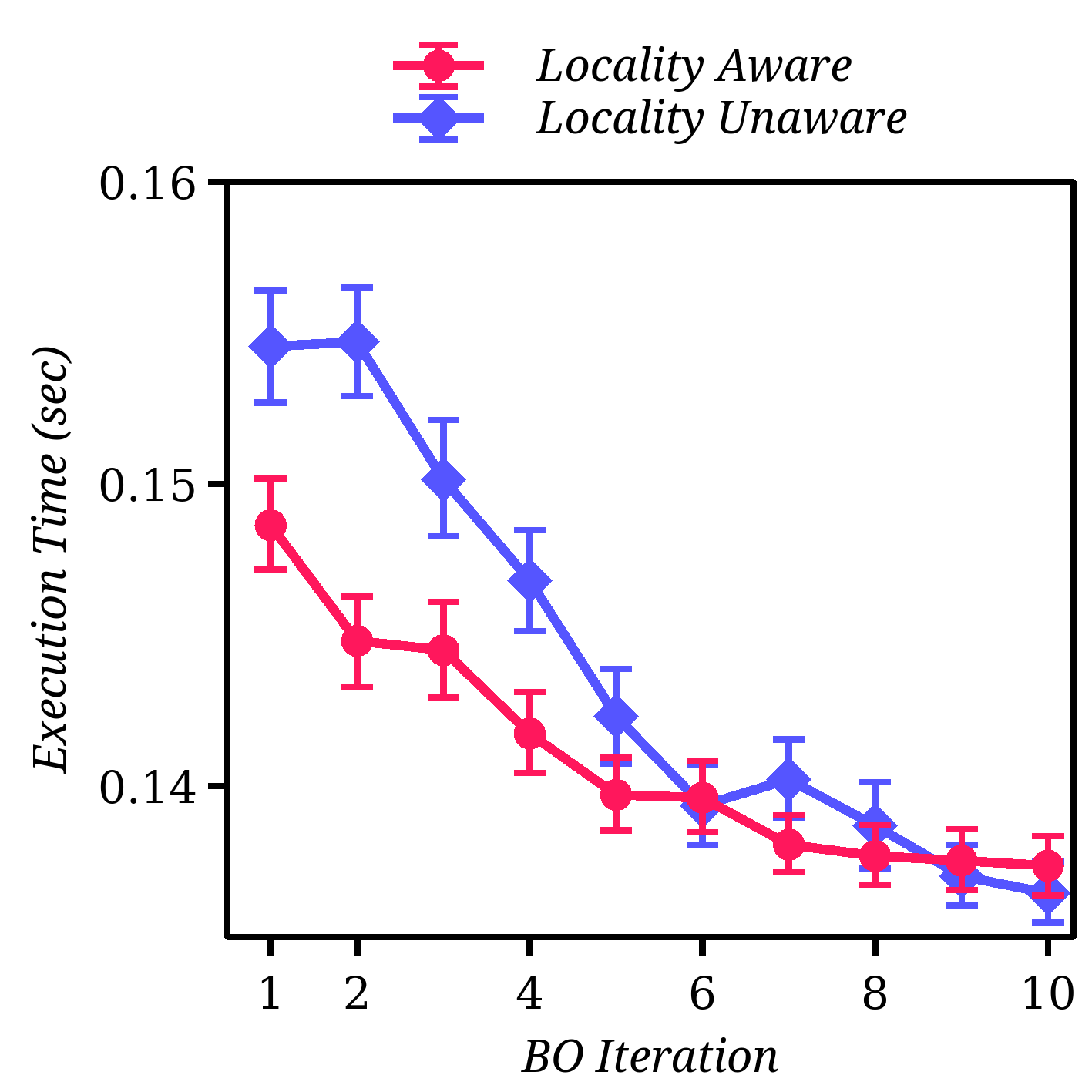}
  \vspace{-0.1in}
  \caption{Convergence plot of the locality-unaware GP and the locality-aware GP on the \code{skitter} workload.
    We can see the execution time decreasing as we run BO.
  We ran BO 30 times with 10 iterations each, and computed the 95\% boostrap confidence intervals.}\label{fig:convergence}
\end{figure}

\begin{figure*}[h]
  \vspace{-0.1in}
  \centering
  \includegraphics[scale=0.45]{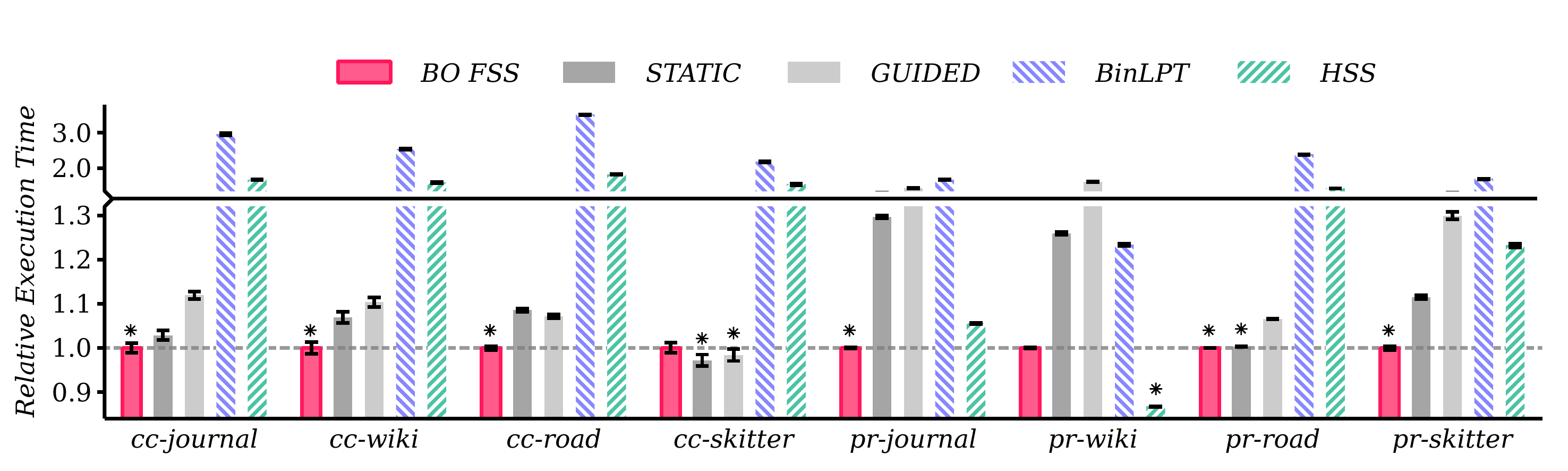}
  \vspace{-0.1in}
  \caption{Execution time comparison of BO FSS against workload-aware methods. We estimate the mean execution time from 256 executions. The error bars show the 95\% bootstrap confidence intervals. The results are normalized by the mean execution time of BO FSS. The methods with the lowest execution time are marked with a star (*). Methods \textit{not} significantly different with the best performing method are also marked with a star (Wilcoxon signed rank test, 1\% null-hypothesis rejection threshold).}\label{fig:high_imbalance}
\end{figure*}


\begin{figure}[t]
  \centering
  \includegraphics[scale=0.4]{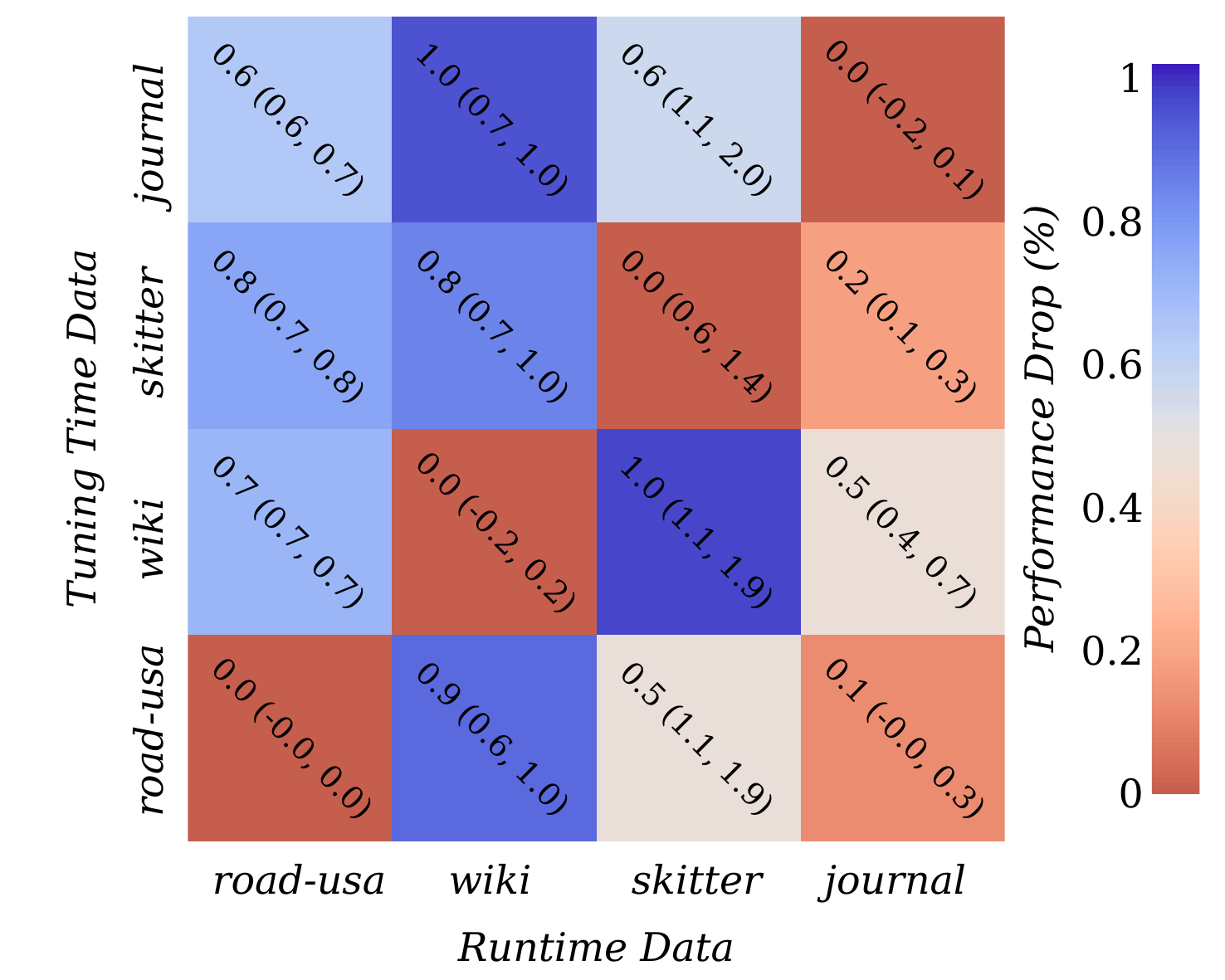}
  \vspace{-0.1in}
  \caption{Effect of mismatching the data used for tuning BO FSS and the data used for execution.
    The rows are the data used for tuning of BO FSS, while the columns are the data used for execution.
    The numbers represent the percentage slowdown relative to the matched case.
    \textcolor{neonblue}{Colder colors} represent more slowdown (\textcolor{red}{hotter} the better).
  }\label{fig:data_cross}
\end{figure}

\vspace{0.05in}

\noindent\textbf{Performance degradation on \code{srad v1}.}
Interestingly, BO FSS does not perform well on two workloads: \code{srad v1} and \code{cc-skitter}.
While the performance difference in \code{cc-skitter} is marginal, the difference in \code{srad v1} is not.
This phenomenon is due to the large deviations in the execution time measurements as shown in Fig.~\ref{fig:outlier}.
That is, large outliers near \(\theta = 0.4\) and \(\theta = 0.8\) deviated the GP predictions (\textcolor{neongreen}{green line}).
Since GPs assume the noise to be Gaussian, they are not well suited for this kind of workload.
A possible remedy is to use \textit{Student-T processes}~\cite{martinez-cantin_practical_2017, shah_bayesian_2013}, shown with the \textcolor{blue}{blue line}.
In Fig.~\ref{fig:outlier}, the Student-T process is much less affected by outliers, resulting in a tighter fit.
Nonetheless, GPs worked consistently well on other workloads.

\vspace{0.1in}

\begin{figure}[t]
  \vspace{-0.1in}
  \centering
  \includegraphics[scale=0.40]{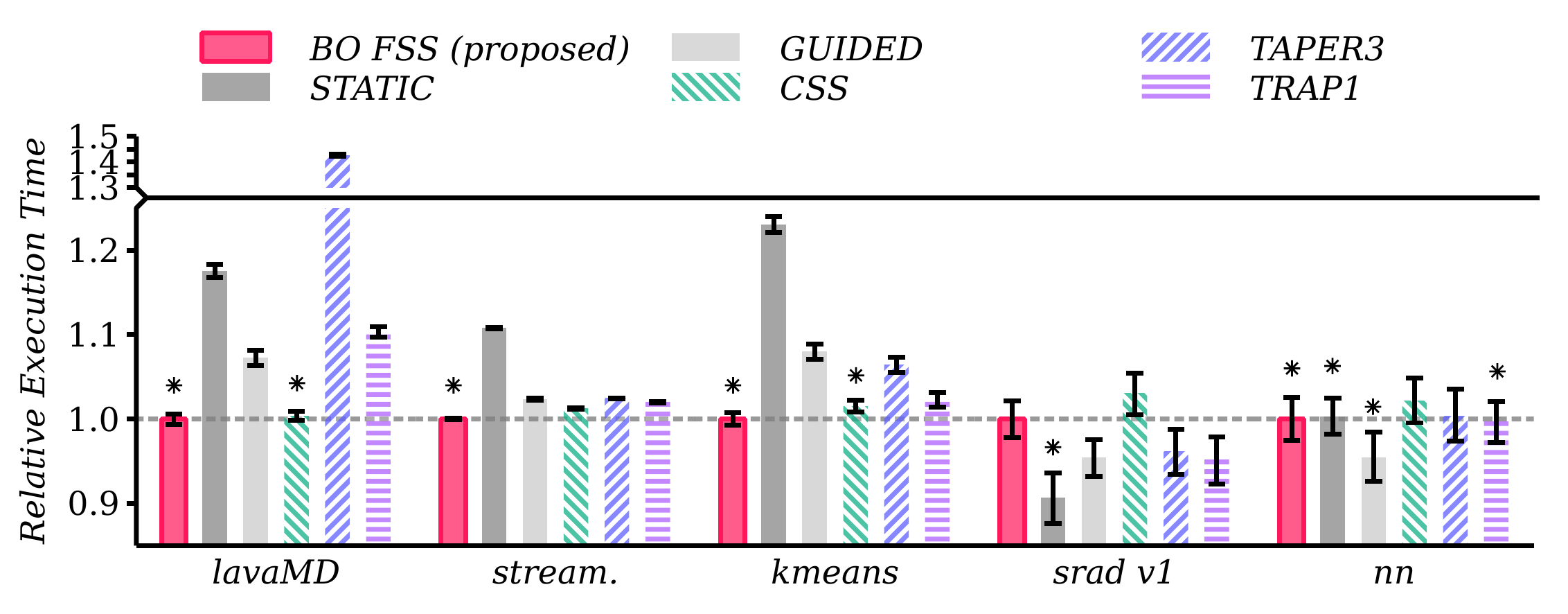}
  \vspace{-0.1in}
  \caption{Execution time comparison of BO FSS against dynamic scheduling methods. We estimate the mean execution time from 256 executions. The error bars show the 95\% bootstrap confidence intervals. The results are normalized by the mean execution time of BO FSS. The methods with the lowest execution time are marked with a star (*). Methods \textit{not} significantly different with the best performing method are also marked with a star (Wilcoxon signed rank test, 1\% null-hypothesis rejection threshold).}\label{fig:low_imbalance}
  \vspace{-0.2in}
\end{figure}

\noindent\textbf{Comparison of Gaussian Process Models.}
We now compare the simple GP construction in Section~\ref{total} and the locality-aware GP construction in Section~\ref{labo}.
We equip BO with each of the models, and run the autotuning process beginning to end 30 times.
The convergence results are shown in Fig.~\ref{fig:convergence}.
We can see that the locality-aware construction converges much quickly.
Note that the shown results are averages.
In the individual results, there are cases where the locality-unaware version completely fails to converge within a given budget.
We thus suggest to use the locality-aware construction whenever possible.
It achieves consistent results at the expense of additional computation during tuning.

\vspace{0.05in}

\noindent\textbf{Remarks.}
Apart from \code{srad v1}, BO FSS performs better than FSS and FAC2 on most workloads.
This indicates that the Gaussian assumption works fairly well in most cases.
We can conclude that our BO augmentation improves the performance of FSS on both workloads with high and low static imbalances.
Our interest is now to see how this improvement compares against other scheduling algorithms.

\vspace{0.05in}

\subsection{Evaluation on Workloads Without Static Imbalance}\label{perf_rodinia}
This section compares the performance of BO FSS against dynamic scheduling methods on workloads where a workload-profile is unavailable or uniformative.
The benchmark results are shown in Fig.~\ref{fig:low_imbalance}.
Out of the 5 workloads considered, BO FSS outperforms all other methods on 3 out of 5 workloads.
On the \code{nn} workload, the difference between all methods is insignificant.
As discussed in Section~\ref{perf_aug}, BO FSS performs poorly on the \code{srad v1} workload.
Note that the same tuning results are used both for Section~\ref{perf_aug} and this experiment.

\vspace{0.05in}

\noindent\textbf{Remarks.}
Compared to other dynamic scheduling methods, BO FSS achieves more consistent performance.
However, because of the turbulence in the tuning process, BO FSS performs poorly on \code{srad v1}.
It is thus important to ensure that BO FSS correctly converges to a critical point before applying it.

\vspace{-0.05in}

\subsection{Evaluation on Workloads With Static Imbalance}\label{perf_gapbs}
This section evaluates the performance of BO FSS against workload-aware methods using workloads with a workload-profile.
The evaluation results are shown in Fig.~\ref{fig:high_imbalance}.
Except for the \code{pr-wiki} workload, BO FSS dominates all considered baselines.
Because of the large number of tasks, both the HSS and BinLPT algorithms do not perform well on these workloads.
Meanwhile, the STATIC and GUIDED strategies are very inconsistent in terms of performance.
On the \code{pr-wiki} and \code{pr-journal} workloads, both methods are nearly 30\% slower than BO FSS.
This means that these algorithms lack workload-robustness unlike BO FSS.

On the \code{pr-wiki} workload which has the most extreme level of static imbalance, HSS performs significantly better.
As discussed in Section~\ref{motivation}, HSS has a very large critical section, resulting in a large amount of scheduling overhead.
However, on the \code{pr-wiki} workload, the inefficiency caused by load imbalance is so extreme compared to the inefficiency caused by the scheduling overhead, giving HSS a relative advantage.

\vspace{0.05in}

\noindent\textbf{Does the input data affect performance?}
BO FSS's performance is tightly related to the individual property of each workload.
It is thus interesting to ask how much the input data of the workload affects the behavior of BO FSS.\
To analyze this, we interchange the data used to tune BO FSS and the data used to measure the performance.
If the input data plays an important role, the discrepancy between the \textit{tuning time data} and the \textit{runtime data} would degrade the performance.
The corresponding results are shown in Fig.~\ref{fig:data_cross} where the entries are the percentage increase in execution time relative to the \textit{matched case}.
Each row represents the dataset used for tuning, while each column represents the dataset used for execution.
The anti-diagonal (bottom left to top right) is the case when the data is matched.
The maximum amount of degradation is caused when we use \code{skitter} for tuning and \code{wiki} during runtime.
Also, the case of using \code{journal} for tuning and \code{wiki} during runtime significantly degrades the performance.
Overall, the \code{wiki} and \code{road} datasets turned out to be the pickiest about the match.
Since both \code{wiki} and \code{road} resulted in high degradation, the amount of imbalance in the data does not determine how important the match is.
However, judging from the fact that the degradation is at most 1\%, we can conclude that BO FSS is more sensitive to the workload's algorithm rather than its input data.

\vspace{0.05in}

\noindent\textbf{Remarks.}
Compared to the workload-aware methods, BO FSS performed the best except for one workload which has the most amount of imbalance.
Excluding this extreme case, the performance benefits of BO FSS is quite large.
We also evaluated the sensitivity of BO FSS on perturbations to the workload.
Results show that BO FSS is not affected much by changes in the input data of the workload.

\subsection{Discussions}
\noindent\textbf{Analysis of overhead.}
BO FSS has specific duties, both online and offline.
When online, BO FSS loads the precomputed scheduling parameter $\theta_i$, measures the loop execution time and stores the pair $(\theta_i,  \tau_i)$ in the dataset $\mathcal{D}_t$.
A storage memory overhead of $O(T)$, where $T$ is the number of BO iterations, is required to store $\mathcal{D}_t$.
This is normally much less than the $O(N)$ memory requirement, where $N$ is the number of tasks, imposed by workload-aware methods.
When offline, BO FSS runs BO using the dataset $\mathcal{D}_t$ and determines the next scheduling parameter $\theta_{i+1}$.
Because most of the actual work is performed offline, the online overhead of BO FSS is almost identical to that of FSS.\
The offline step is relatively expensive due to the computation complexity of GPs.\
Fortunately, BO FSS converges within 10 to 20 iterations for most cases. This allows the computational cost to stay within a reasonable range.

\vspace{0.05in}
\noindent\textbf{Limitation.}
When the target loop is not to be executed for a significant amount of time, BO FSS does provide significant benefits, as it requires time for offline tuning.
However, HPC workloads are often long-running and reused over time. 
For this reason, BO FSS should be applicable for many HPC workloads.

\vspace{0.05in}

\noindent\textbf{Portability.}
%
When solving the optimization problem in~\eqref{eq:opt_problem} with BO, the target system becomes part of the objective function.
As a result, BO FSS automatically takes into account the properties of the target system.
This fact makes BO FSS highly portable.
At the same time, as the experimental results of Fig.~\ref{fig:data_cross} imply, instead of directly operating on the full target workload, it should be possible to use much cheaper proxy workloads for tuning BO FSS.


\section{Related Works}\label{related}
\noindent\textbf{Classical dynamic loop scheduling methods.}
To improve the efficiency of dynamic scheduling, many classical algorithms are introduced such as CSS~\cite{kruskal_allocating_1985}, FSS~\cite{hummel_factoring_1992}, TSS~\cite{tzen_trapezoid_1993}, BOLD~\cite{hagerup_allocating_1997}, TAPER~\cite{lucco_dynamic_1992} and BAL~\cite{bast_scheduling_2000}. However, most of these classic algorithms are derived in a limited theoretical context with strict statistical assumptions.
Such an example is the \(i.i.d.\) assumption imposed on the workload.

\vspace{0.05in}

\noindent\textbf{Adaptive and workload-aware methods.}
To resolve this limitation, adaptive methods are developed starting from the \textit{adaptive FSS} algorithm~\cite{banicescu_load_2002}.
Recently, workload-aware methods including HSS~\cite{kejariwal_historyaware_2006} and BinLPT~\cite{penna_binlpt_2017, penna_comprehensive_2019} are introduced.
These scheduling algorithms explicitly require a workload-profile before execution and exploit this knowledge in the scheduling process.
On the flip side, this requirement makes these methods difficult to use in practice since the exact workload-profile may not always be available beforehand.
In contrast, our proposed method is more convenient since we only need to measure the execution time of a loop.
Also, the overall concept of our method is more flexible; it is possible to plug in our framework to any parameterized scheduling algorithm, directly improving its robustness.

\vspace{0.05in}

\noindent\textbf{Machine learning based approaches.}
Machine learning has yet to see many applications in parallel loop scheduling.
In~\cite{wang_mapping_2009}, Wang and O'Boyle use compiler generated features to train classifiers that select the best-suited scheduling strategy for a workload.
This approach contrasts with ours since it does not improve the effectiveness of the chosen scheduling algorithm.
On the other hand, Khatami et al.\ in~\cite{khatami_hpx_2017} recently used a logistic regression model for predicting the optimal chunk size for a scheduling strategy, combining CSS and work-stealing.
Similarly, Laberge et al.~\cite{laberge_scheduling_2019a} propose a machine-learning based strategy for accelerating linear algebra applications.
These supervised-learning based approaches are limited in the sense that they are not yet well understood: their performance is dependent on the quality of the training data.
It is unknown how well these approaches \textit{generalize} across workloads from different application domains.
In fact, quantifying and improving generalization is still a central problem in supervised learning.
Our method is free of these issues since we directly optimize the performance for a target workload.


\section{Conclusion}\label{conclusion}
In this paper, we have presented BO FSS, a data-driven, adaptive loop scheduling algorithm based on BO.
The proposed approach automatically tunes its performance to the workload using execution time measurements.
Also, unlike the scheduling algorithms that are inapplicable to some workloads, our approach is generally applicable.
We implemented our method on the \code{OpenMP} framework and quantified its performance as well as its robustness on realistic workloads.
BO FSS has consistently performed well on a wide range of real workloads, showing that it is robust compared to other loop scheduling algorithms.
Our approach motivates the development of computer systems that can automatically adapt to the target workload.

At the moment, BO FSS assumes that the properties of the workload do not change during execution.
For this reason, BO FSS does not address some crucial scientific workloads, such as adaptive mesh refinement methods.
These types of workloads dynamically change during execution, depending on the computation results.
It would be interesting to investigate automatic tuning-based scheduling algorithms that can target such types of workloads in the future.

 \begin{table*}[h]
   \footnotesize
  \caption{Implementation Details of Considered Baselines}\label{table:params}
 \centering
    \begin{tabular}{ccc}
     \toprule
   \textbf{Type}  & \textbf{Chunk Size Equation} & \textbf{Parameter Setting} \\
     \midrule
 CSS~\cite{kruskal_allocating_1985} &   
 $\begin{aligned}
K =  {\left(\frac{h}{\sigma} \frac{\sqrt{2}N}{P \sqrt{\log P}}\right)}^{2/3}
\end{aligned}$
&
$h$, $\sigma$, $\mu$ (measured values)

\\

\cmidrule(l r){1-3}
 TAPER~\cite{lucco_dynamic_1992} &
$\begin{aligned}
v_\alpha &= \alpha \frac{\sigma}{\mu}, \;\; x_i = \frac{R_i}{P} + \frac{K_{\text{min}}}{2}, \;\; R_{i+1} = R_i - K_i \\
K_i &= \max(K_{\text{min}}, \, x_i + \frac{v_a^2}{2} - v_\alpha \sqrt{2 x_i + \frac{v_\alpha^2}{4} })
\end{aligned}$
&
$\begin{aligned}
v_\alpha = 3, K_{\text{min}} = 1
\end{aligned}$
\\

\cmidrule(l r){1-3}
TSS~\cite{tzen_trapezoid_1993} &
$\begin{aligned}
&\delta = \frac{K_f - K_l}{N - 1}, \;\; K_0 = K_f \\
&K_{i+1} = \max(K_i - \delta, K_l)
\end{aligned}$
&
$\begin{aligned}
K_f = \frac{N}{2P}, \;\; K_{l} = 1, 
\end{aligned}$

\\
    \bottomrule
    \end{tabular}
  \end{table*}

\ifCLASSOPTIONcompsoc
\section*{Acknowledgments}
 \else
  \section*{Acknowledgment}
\fi

The authors would like to thank the reviewers for providing precious comments enriching our work, Pedro Henrique Penna for the helpful discussions about the BinLPT scheduling algorithm, Myoung Suk Kim for his insightful comments about our statistical analysis and Rover Root for his helpful comments about the scientific workloads considered in this work.%

\ifCLASSOPTIONcaptionsoff
  \newpage
\fi



\bibliographystyle{IEEEtran}
\bibliography{bstcontrol,references}

%

%

\vspace{-0.4in}

\begin{IEEEbiography}[{\includegraphics[width=1in,height=1.25in,clip,keepaspectratio]{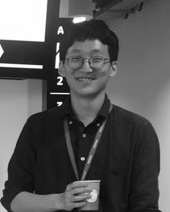}}]{Kyurae~Kim}
(Student Member, IEEE), is working towards his B.S. degree with the Department of Electronics Engineering, Sogang University, Seoul, South Korea.

  His research interests lie in the duality of machine learning and computer systems, including parallel computing, compiler runtime environments, probabilistic machine learning and Bayesian inference methods.
\end{IEEEbiography}

\vspace{-0.4in}

\begin{IEEEbiography}[{\includegraphics[width=1in,height=1.25in,clip,keepaspectratio]{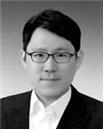}}]{Youngjae~Kim}
  (Member, IEEE) received the B.S. degree in computer science from Sogang University, South Korea, in 2001, the MS degree in computer science from KAIST, in 2003, and the PhD degree in computer science and engineering from Pennsylvania State University, University Park, Pennsylvania, in 2009.

  He is currently an associate professor with the Department of Computer Science and Engineering, Sogang University, Seoul, South Korea.
  Before joining Sogang University, Seoul, South Korea, he was a R\&D staff scientist with the US Department of Energy’s Oak Ridge National Laboratory (2009--2015) and as an assistant professor at Ajou University, Suwon, South Korea (2015--2016).
  His research interests include operating systems, file and storage systems, parallel and distributed systems, computer systems security, and performance evaluation.
\end{IEEEbiography}

\vspace{-0.4in}

\begin{IEEEbiography}[{\includegraphics[width=1in,height=1.25in,clip,keepaspectratio]{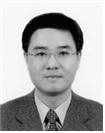}}]{Sungyong~Park}
  (Member, IEEE) received the B.S. degree in computer science from Sogang University, Seoul, South Korea, and both the MS and PhD degrees in computer science from Syracuse University, Syracuse, New York.

  He is currently a professor with the Department of Computer Science and Engineering, Sogang University, Seoul, South Korea.
  From 1987 to 1992, he worked for LG Electronics, South Korea, as a research engineer.
  From 1998 to 1999, he was a research scientist at Bellcore, where he developed network management software for optical switches.
  His research interests include cloud computing and systems, high performance I/O and storage systems, parallel and distributed system, and embedded system.
\end{IEEEbiography}




\end{document}